\documentclass[fleqn,usenatbib]{mnras}

\usepackage[T1]{fontenc}
\usepackage{ae,aecompl}

\usepackage{graphicx}	% Including figure files
\usepackage{amsmath}	% Advanced maths commands
\usepackage{amssymb}	% Extra maths symbols
\usepackage{float}
\usepackage{natbib}
\usepackage{morefloats}
\usepackage{dcolumn}
\usepackage{amsmath}
\usepackage{latexsym}
\usepackage{longtable}
\usepackage{lipsum} 
\title[SN 2014dt]{Exploring the optical behaviour of a type Iax supernova SN 2014dt}
\author[Mridweeka Singh et al.]{
Mridweeka Singh$^{1,4}$\thanks{E-mail: mridweeka@aries.res.in, yashasvi04@gmail.com},
Kuntal Misra$^{1}$,
D. K. Sahu$^{2}$,
Raya Dastidar$^{1,5}$,
\newauthor
Anjasha Gangopadhyay$^{1,4}$,
Subhash Bose$^{3}$,
Shubham Srivastav$^{2}$,
G. C. Anupama$^{2}$,
\newauthor
N. K. Chakradhari$^{4}$,
Brajesh Kumar$^{2}$,
Brijesh Kumar$^{1}$,
S. B. Pandey$^{1}$
\\
$^{1}$Aryabhatta Research Institute of observational sciencES, Manora Peak, Nainital 263 001, India\\
$^{2}$Indian Institute of Astrophysics, Koramangala, Bangalore 560 034, India\\
$^{3}$Kavli Institute for Astronomy and Astrophysics, Peking University, 5 Yiheyuan Road, Haidian District, Beijing 100871, P.R. China\\
$^{4}$School of Studies in Physics and Astrophysics, Pandit Ravishankar Shukla University, Chattisgarh 492 010, India \\
$^{5}$Department of Physics \& Astrophysics, University of Delhi, Delhi-110 007, India\\
}

\date{Accepted XXX. Received YYY; in original form ZZZ}

\pubyear{2016}

\begin{document}
\label{firstpage}
\pagerange{\pageref{firstpage}--\pageref{lastpage}}
\maketitle

\begin{abstract}

We present optical photometric (upto $\sim$410 days since $B$$_{max}$) and spectroscopic (upto $\sim$157 days since $B$$_{max}$) observations of a Type Iax supernova (SN) 2014dt located in M61. SN 2014dt is one of the brightest and closest (D $\sim$ 20 Mpc) discovered Type Iax SN. SN 2014dt best matches the light curve evolution of SN 2005hk and reaches a peak magnitude of $M$$_B$ $\sim$-18.13$\pm$0.04 mag with $\Delta m_{15}$ $\sim$1.35$\pm 0.06$ mag. The early spectra of SN 2014dt are similar to other Type Iax SNe, whereas the nebular spectrum at 157 days is dominated by narrow emission features with less blending as compared to SNe 2008ge and 2012Z. The ejecta velocities are between 5000 to 1000 km sec$^{-1}$ which also confirms the low energy budget of Type Iax SN 2014dt as compared to normal Type Ia SNe. Using the peak bolometric luminosity of SN 2005hk we estimate $^{56}$Ni mass of $\sim$0.14 M$_{\odot}$ and the striking similarity between SN 2014dt and SN 2005hk implies that a comparable amount of $^{56}$Ni would have been synthesized in the explosion of SN 2014dt.
\end{abstract}


%%%%%%%%%%%%%%%%%%%%%%%%%%%%%%%%%%%%%%%%%%%%%%%%%%

%%%%%%%%%%%%%%%%% BODY OF PAPER %%%%%%%%%%%%%%%%%%

\section{Introduction}

The last two decades have witnessed the advent of a new subclass of Type Ia SNe which are grouped together and are commonly known as Type Iax SNe \citep{2013ApJ...767...57F}. The rate of occurrence of Type Iax SNe is 31$^{+19}_{-13}$ of every normal 100 Type Ia SNe \citep{2013ApJ...767...57F}. Type Iax SNe exhibit distinct properties both photometrically and spectroscopically when compared to the normal Type Ia SNe. The secondary peak in NIR light curves of Type Ia SNe is not seen in the NIR light curves of Type Iax SNe \citep{2003PASP..115..453L}. A wide range is seen in peak absolute brightness ($M$${_V}$ = -14 to -18 mag). Their expansion velocities are ($\sim$4000 to $\sim$9000 km s$^{-1}$) half of that of Type Ia SNe ($\sim$10,000 to $\sim$15,000 km s$^{-1}$). For a small sample of SN 2002cx like objects \cite{2010ApJ...720..704M} suggested that there were correlations between peak luminosity, light curve shape and ejecta velocity. \cite{2011ApJ...731L..11N} also derived scaling relations between ejecta velocity and decline rate using Arnett's formulation. It was found by \cite{2011ApJ...731L..11N} that the ejecta mass of Type Iax SNe was between 1-1.4 M{$_\odot$} (SN 2008ha being an exception). With a slightly larger sample, the ejecta mass of Type Iax SNe was estimated to be 0.5$\pm$0.2 M{$_\odot$} \citep{2013ApJ...767...57F}. With more number of Type Iax discovered, correlations between decline rate and ejecta velocity, peak luminosity and decline rate were found by \cite{2013ApJ...767...57F} and \cite{2016A&A...589A..89M}. They are sub-luminous events but their early time spectra are similar to SN 1991T-like over-luminous events. Often their maximum light spectra show signs of unburned carbon, $\sim$82$\%$ of SNe Iax have clear absorption of carbon whereas only 30$\%$ of SNe Ia have carbon in their pre-maximum spectra \citep{2013ApJ...767...57F}. The nebular spectra of Type Ia SNe are associated with forbidden emission lines of iron group elements whereas Type Iax have permitted lines of these iron group elements along with intermediate-mass elements (S, Ca etc.). It is also seen that the late-time spectra of Type Iax SNe have calcium interior to iron \citep{2013ApJ...767...57F} which is opposite to what we see in Type Ia SNe. The two component model proposed by \cite{2016MNRAS.461..433F} discusses the origin of broad emission lines from the ejecta and narrow forbidden lines originating from a wind which is thought to be associated to the remnant of the progenitor.

The progenitors are a very good source for a clear distinction between various subclasses of Type Ia SNe. \cite{2015ApJ...808..138L} proposed different explosion scenario and progenitor systems to investigate the nature of these event. It is widely accepted that thermonuclear explosion of carbon oxygen white dwarf (Chandrasekhar or sub-Chandrasekhar-mass limit of white dwarf) gives rise to the peculiar Type Iax SNe \citep{2009AJ....138..376F,2012ApJ...761L..23J,2013MNRAS.429.2287K,2014MNRAS.438.1762F,2015A&A...573A...2S,2015MNRAS.450.3045K,2015A&A...574A..12L}. But one of the Type Iax SNe, SN 2008ha supports the core-collapse scenario with the progenitor being a hydrogen deficient massive star having main sequence mass of 13 M$_{\odot}$ \citep{2009Natur.459..674V,2010ApJ...719.1445M}. \cite{2009AJ....138..376F,2010ApJ...708L..61F} presented maximum light spectrum of SN 2008ha with association of C/O burning features which is in contradiction with core collapse origin. One of the most important difference between Type Ia and Type Iax SNe is that there is no direct observational evidence of a progenitor system of Type Ia SNe yet. In contrary to this probable progenitors are imaged for Type Iax SNe. In the case of SN 2012Z, a coincident luminous blue source at the location of the SN was detected in the pre-explosion {\it HST} images \citep{2014Natur.512...54M}. Other two SNe with pre explosion images are SNe 2008ge and 2014dt (for details see \cite{2010AJ....140.1321F,2015ApJ...798L..37F}). SNe 2004cs and 2007J are two special cases in which the classification is debated. They were classified both as Type IIb and Iax showing He features by several authors (for details see \cite{2005PASP..117..132R,2007CBET..809....1L,2007CBET..817....1F,2007CBET..926....1F,2009AJ....138..376F,2013ApJ...767...57F,2015ApJ...799...52W}). The claim of helium connection with the rare Type Iax SNe and hence their progenitors is questionable which is solely based on two objects \citep{2015ApJ...799...52W}.
     
In this paper we examine the photometric and spectroscopic properties of a peculiar Type Iax SN 2014dt. In section 3 we describe the data acquisition and reduction procedure. In section 4 we discuss the distance estimates and extinction of the host galaxy. The estimated magnitudes are used to construct and study the multi band light curves in section 5. Section 5 also discusses the absolute magnitude, colour evolution, and pseudo bolometric light curves of SN 2014dt along with other Type Iax SNe. A detailed description of the spectral evolution and spectral modelling is discussed in section 6. In section 7 we summarize the results obtained.
 
\section{SN 2014dt}  
SN 2014dt was discovered on 2014 October 29.838 UT (JD = 2456960.338) in an unfiltered CCD frame with a 0.50m f/6.8 reflector telescope at Takanezawa, Tochigi-kenn in the galaxy M61 \citep{2014CBET.4011....1N}. On October 30.50 UT, the SN was 13.6 mag bright in $V$ filter. SN 2014dt is one of the brightest (peak magnitude $\sim$ 13.6 in $V$ filter) and closest discovered Type Iax SN (D $\sim$ 20 Mpc). The SN is 33$''$.9 east and 7$''$.2 south of the centre of the host galaxy M61 which has hosted six other Type II SNe. Spectroscopic observations of SN 2014dt with 182 cm Copernico Telescope on October 31.20 UT showed characteristics of a peculiar Type Ia SN with a relatively blue continuum, weak Si II absorption at 6350 \AA~and Fe II lines at 4300 \AA~and 5000 \AA~ \citep{2014CBET.4011....2O} similar to SN 2002cx at +17 days classifying it as a Type Iax SN. Along with SNe 2008ge and 2012Z, SN 2014dt is the third member of the Type Iax class with available pre-explosion images. Based on the pre-explosion images \cite{2015ApJ...798L..37F} suggested that progenitor of SN 2014dt could be a carbon oxygen white dwarf plus Helium star system with a hotter donor. They suggest that the progenitor system of SN 2014dt may have a low-mass red giant or a main sequence star as a companion (more details are given in section 6). The details of the host galaxy M61 and SN 2014dt are given in Table \ref{tab:sn14dt_m61_detail}.

\begin{figure}
	\begin{center}
		\includegraphics[scale=0.3]{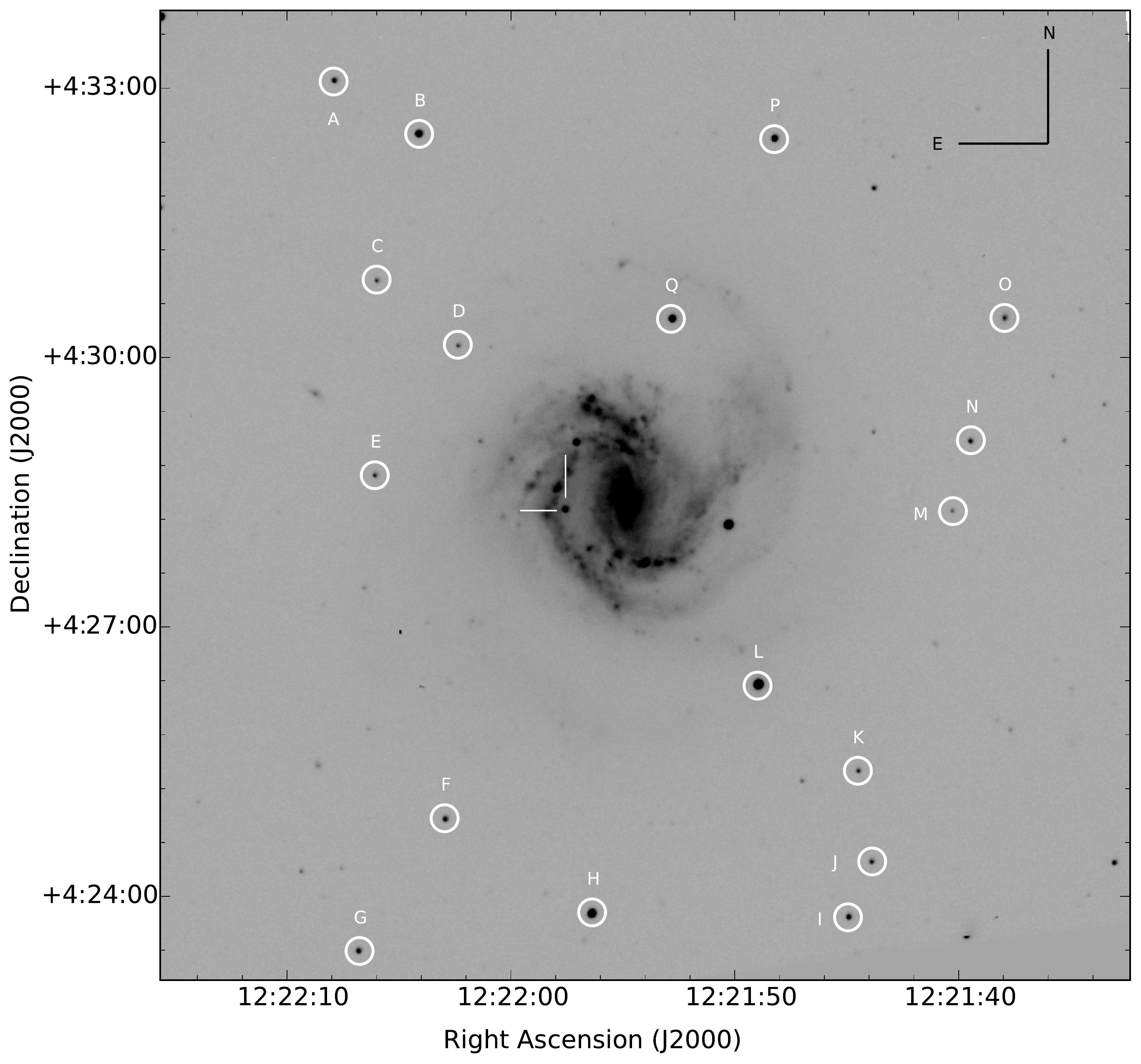}
	\end{center}
	\caption{SN 2014dt in M61, star ID for local standards have been marked. This image is taken on 20 January, 2015 in $V$ band from 104 cm ST which covers an area of about 13.5$\times$13.5 arcmin$^2$. }
	\label{fig:id_chart}
\end{figure} 
\section{Data acquisition and Reduction}
\subsection{Optical Photometric Observations}
A rigorous follow up campaign of SN 2014dt started immediately after the discovery with 104 cm Sampurnanand Telescope (ST) \citep{1999CSci...77..643G}, 130 cm Devasthal Fast Optical Telescope (DFOT) \citep{2012SPIE.8444E..1TS} situated in ARIES, Nainital (India) and 201 cm Himalayan Chandra Telescope (HCT) \cite{2010ASInC...1..193P} and references therein) situated in Indian Astronomical Observatory (IAO) Hanle (India). The details of the detectors used with different telescopes are given in Table \ref{tab:details_instrument_detectors}. The photometric observations of SN 2014dt started $\sim$1 day after discovery and lasted upto $\sim$400 days. Along with the SN field, bias frames for removing zero integration noise and flat frames for removing non uniformity of pixel to pixel response were also taken. The science frames were corrected for bias and flat fields by using the packages available in IRAF \footnote{IRAF
stands for Image Reduction and Analysis Facility distributed by the National Optical Astronomy Observatories which is operated by the Association of Universities for research in Astronomy, Inc., under cooperative agreement with the National Science Foundation.}. Cosmic rays were removed by using the task L.A. Cosmic \citep{2001PASP..113.1420V}. To improve S/N ratio wherever multiple frames were available on a single night, images were co-added. Since the SN is situated away from the galaxy centre, near one of the spiral arms of the galaxy and lies in a relatively clean part of the galaxy, we did not perform template subtraction to remove the galaxy contribution. We performed PSF photometry using packages in DAOPHOT II \citep{1987PASP...99..191S}  to estimate the SN magnitudes. We took the first aperture equal to mean FWHM of the frame and inner sky annulus was taken to be four times the FWHM. An eight pixel wide sky annulus was selected with respect to inner sky annulus.

To convert the instrumental magnitudes into standard magnitudes we observed the Landolt equatorial standards  \citep{2009AJ....137.4186L} on 24 November 2014 and 01 December 2015 along with the SN field in $VRI$ and {\it UBVRI} bands, respectively. These observations were done under good photometric conditions (seeing in $V$ band was $\sim$1.7 arcsec). We selected stars associated with PG 0231, PG 0918, PG 0942, PG 2213, PG 2331, SA 92 and SA 95 fields for observations. Their brightness range varied between 16.28~\textless~$V$ \textless~12.27 and colour range was -0.32 \textless~${\it B}-{\it V}$ \textless~1.45 mag. The standard fields were observed at different airmass ranging between 1.9 to 1.3 on 01 December 2015. We used the standard magnitudes and the instrumental magnitudes of the Landolt field stars to simultaneously fit for the extinction coefficient, colour coefficient and zero point following the least square regression technique described in \cite{1992JRASC..86...71S}. However, Landolt standards on 24 November 2014 were observed only for estimating the zero points and we used site extinction values \citep{2008BASI...36..111S} for applying the extinction correction while extinction values were derived for the data acquired on 01 December 2015. 

The fitted values of these coefficients were used to transform the instrumental magnitudes of the standard stars to standard magnitudes. The root-mean-squared (rms) scatter between transformed and standard magnitude of Landolt stars was found to be $\sim$0.04 mag in $B$ band, $\sim$0.02 mag in $V$ band, $\sim$0.02 mag in $R$ band and $\sim$0.03 mag in $I$ band. Using these transformation equations, we generated 17 local standards in the SN field which were non-variables. Final magnitudes (weighted average for two sets of observations for $VRI$ bands) of these 17 local standards are listed in Table \ref{tab:standard_star_table} and marked in Figure \ref{fig:id_chart} along with the SN location. For all other nights zero points were determined and SN magnitudes were calibrated differentially using these local standards. The errors due to calibration and photometry were added in quadrature to estimate the final error in SN magnitudes. The calibrated SN magnitudes and the associated errors in {\it BVRI} filters are listed in Table \ref{tab:photometric_observational_log}. 

\subsection{Optical Spectroscopic Observations}
\label{Spectroscopy}
Long slit low resolution spectroscopic data at 18 epochs were obtained with HCT, equipped with Hanle Faint Object Spectrograph and Camera (HFOSC). To cover the entire optical region, pair of grisms Gr7 (3800 \AA~- 7800 \AA) with a resolution 1330 and Gr8 (5800 \AA~- 9200 \AA) with a resolution 2190 is used with HFOSC for spectroscopic observations. The slit width (0.77 and 1.92 arcsec) was chosen so as to avoid contamination due to host. Exposure time was varied between 600 to 1800 seconds in order to obtain good SNR spectra. For wavelength calibration arc lamps (FeAr and FeNe) were observed just after SN exposure and for flux calibration spectro photometric standard stars (Feige 34, Feige 110 and HZ 44) were observed each night along with the SN. Necessary pre-processing and spectral reduction were done with standard tasks in IRAF. To cross check the wavelength calibration, OI emission lines at 5577, 6300 and 6364 \AA~ were used and in some cases wavelength shift within 0.2 to 4 \AA~was found. This shift was applied to correct the spectrum. The blue and red region spectra in Gr7 and Gr8 were combined by applying a scaling factor to get the final spectrum on a relative flux scale. The relative flux spectra were brought to an absolute flux scale by estimating the scale factors from {\it BVRI} magnitudes. The telluric lines were not removed from the spectra. Spectra were corrected for redshift by using DOPCOR. The log of spectroscopic observations is given in Table \ref{tab:spectroscopic_observations}. 

\section{Distance and Extinction } 
Several distance measurements ranging between 7.59 Mpc  \citep{1984A&AS...56..381B} and 35.50$\pm$0.25 Mpc \citep{1994ApJ...433...19S} are available in NED for the host galaxy of SN 2014dt. A weighted average of recent distance measurements with different methods viz CO and HI \citep{1997A&A...323...14S}, SCM \citep{2011ApJ...736...76R,2015ApJ...799..215P}, EPM \citep{2014ApJ...782...98B} and photospheric magnitude method \citep{2014AJ....148..107R} gives a value 15.56$\pm$0.15 Mpc. \cite{2015ApJ...798L..37F} used a distance of 12.3 Mpc to put constraints on the progenitor of SN 2014dt. On the other hand \cite{2016ApJ...816L..13F} argue that absolute magnitudes of SN 2014dt are similar to absolute magnitudes of SNe 2005hk and 2012Z if a distance of 19.3 Mpc is adopted.  Using the redshift z = 0.005224$\pm$0.000007 \citep{1985AJ.....90.1681B}, we estimate the luminosity distance of M61 to be 21.44$\pm$0.03 Mpc for $\it{H}$$_o$ = 73$\pm$5 km s$^{-1}$ Mpc$^{-1}$, $\Omega$$_m$ = 0.27 and $\Omega$$_v$ = 0.73.  Since there is a range of distance measurements available, we adopt the luminosity distance for further analysis in the present work.

The galactic reddening in the direction of SN 2014dt is {\it E(B-V)} = 0.02 mag \citep{2011ApJ...737..103S}. The blue continuum and absence of Na ID absorption lines in the spectrum of SN 2014dt hints towards a negligible host galaxy reddening which is also supported by the fact that SN 2014dt lies in a relatively clean part of the face-on spiral galaxy M61 \citep{2015ApJ...798L..37F}. \cite{2016ApJ...816L..13F} compared the colour evolution of SN 2014dt with SNe 2005hk and 2012Z (hosts of both of these SNe have low extinction) to constrain the low reddening in SN 2014dt. We therefore use only the galactic reddening value as the total reddening with {\it E(B-V)} = 0.02 mag and {\it R$_{v}$} = 3.1 in our work.         

\section{Temporal evolution of SN 2014dt}
\subsection{Prime light curve features}

\begin{figure}
	\begin{center}
		\includegraphics[scale=0.50]{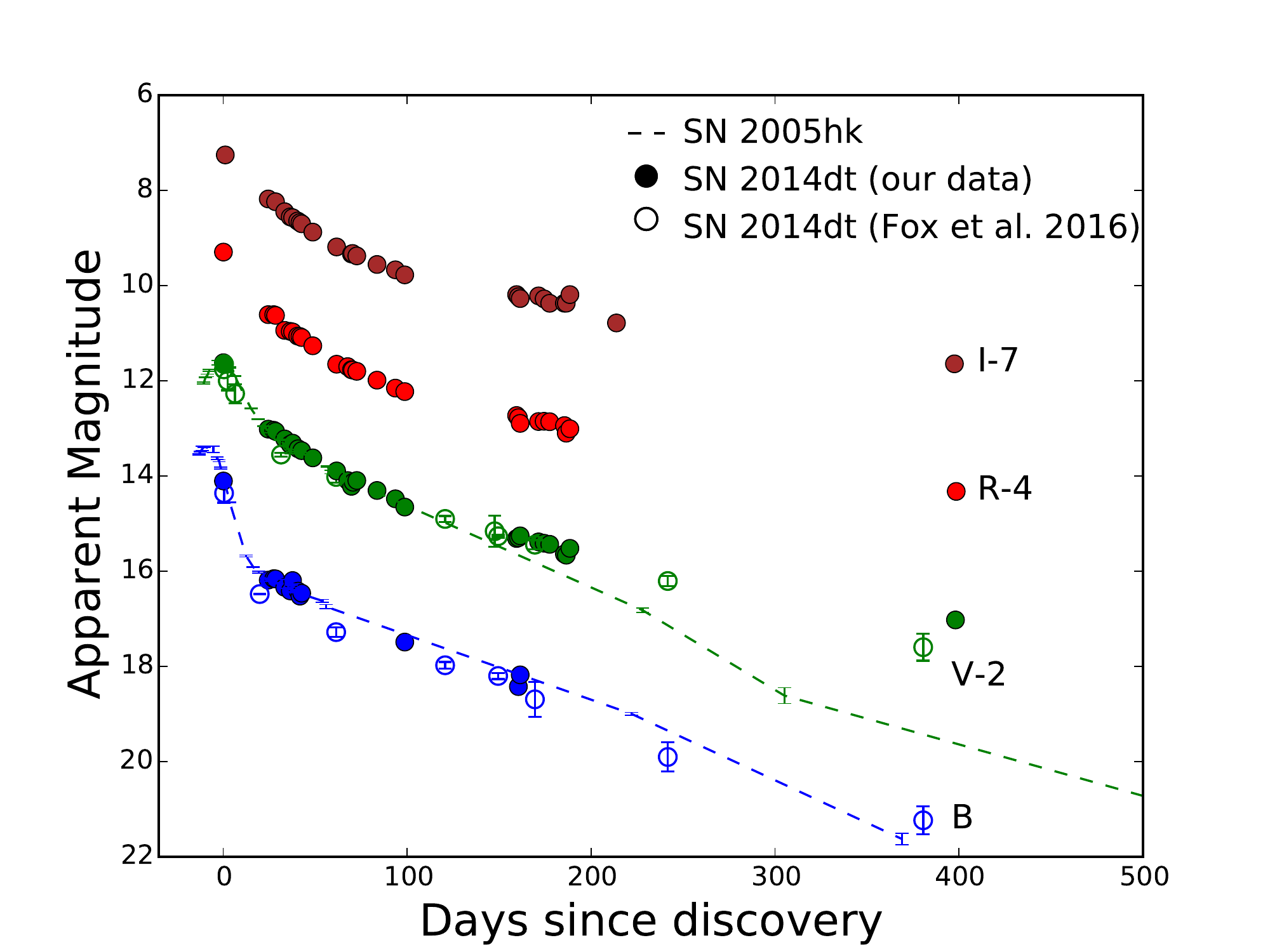}
	\end{center}
	\caption{Broadband {\it BVRI} light curves of SN 2014dt along with best fit templates of SN 2005hk in B and V bands. We have supplemented the data from \citet{2016ApJ...816L..13F} in B and V bands. The light curves in different bands are shifted arbitrarily for clarity. The error bars are smaller than the point size.}
	\label{fig:light_curve_template_plot}
\end{figure}

Figure \ref{fig:light_curve_template_plot} shows the light curve evolution of SN 2014dt in {\it BVRI} bands upto $\sim$ 400 days after discovery. We started the observing campaign $\sim$ 1 day after discovery. The SN was discovered post peak, it is therefore difficult to estimate the peak magnitude and the time of maximum. We adopt template fitting method for estimating magnitudes at maximum in different bands and time of maximum using a $\chi$$^2$ minimization technique which solves simultaneously for both peak magnitude and time of maximum. The template light curves are stretched in time and scaled in magnitudes to get the best match with the observed data set. We take SN 2005hk \citep{2008ApJ...680..580S,2014ApJ...786..134M} and SN 2012Z \citep{2015ApJ...806..191Y} as templates which fall in the category of well observed Type Iax SNe. We supplement our {\it B} and {\it V} band data of SN 2014dt with that published in \cite{2016ApJ...816L..13F} to increase the sampling of our light curve. SN 2005hk provides the best match with SN 2014dt. The {\it B} and {\it V} band light curves of SN 2014dt with best fit templates of SN 2005hk are shown in Figure \ref{fig:light_curve_template_plot}. The best fit estimates the time of $B$$_{max}$ to be JD = 2456950.34. The results obtained by template fitting are listed in Table \ref{tab:decay rate}.  \cite{2016ApJ...816L..13F} also compared the multi-wavelength light curve of SN 2014dt with SNe 2005hk and 2012Z and conclude that SN 2014dt peaked around 2014 October 20 (MJD = 2456950). Using the spectra taken on 1.4 and 19.6 rest frame days after discovery, \cite{2016MNRAS.461..433F} constrain that SN 2014dt was discovered +4$\pm$7 days since maximum and estimate the {\it B}$_{max}$ to be 2014 October 25 (JD = 2456955.7). Our estimation of {\it B}$_{max}$ (JD = 2456950.34) is consistent with the values derived in \cite{2016ApJ...816L..13F} and \cite{2016MNRAS.461..433F}. We use our estimate of {\it B}$_{max}$ (JD = 2456950.34) throughout this paper and consider this as the zero phase in time. Combining our {\it B} band data and that taken from \cite{2016ApJ...816L..13F} we estimate the luminosity decline rate $\Delta$$m{_{15}}$ (change in the magnitude of SN in {\it B} band at maximum and after 15 days of maximum) for SN 2014dt to be 1.35$\pm$0.06 mag which is less than that obtained for SN 2005hk ($\Delta$$m{_{15}}$ = 1.68$\pm$0.05 mag; \cite{2008ApJ...680..580S}). 

Radioactive decay of $^{56}$Ni to $^{56}$Co is responsible for early time light curves of Type Ia SNe and the late-time light curves are powered by decay of $^{56}$Co to $^{56}$Fe. In the case of SN 2014dt light curves follow a linear decline upto 100 days after that a flattening is noticed in the light curves. The decay rates in {\it BVRI} bands between 34 to 108 days are 1.96$\pm$0.07, 2.46$\pm$0.03, 2.43$\pm$0.03 and 2.14$\pm$0.03 mag (100 days)$^{-1}$ respectively. The early time decay rates of SNe 2005hk and 2014dt are listed in Table \ref{tab:decay rate}. Up to 300 days the light curves of normal Type Ia SNe are governed by the trapping of gamma rays after which the light curves transitions to the regime governed by full trapping of positrons. Our optical light curves upto $\sim$100 days are well within gamma ray regime. At very late epochs (298 to 326 days) a flattening is reported in the IR light curves of SN 2014dt by \cite{2016ApJ...816L..13F}. The unavailability of data points between $\sim$ 200 to 400 days in our optical light curves does not allow us to make a similar comparison with the NIR light curves. In the case of SN 2014dt we also estimate the late-time decay rates during 168-200 days since {\it B}$_{max}$ in {\it VRI} bands and are listed in Table \ref{tab:decay rate}. We find that the late-time decay rates of SN 2014dt in {\it V} band is steeper than the standard $^{56}$Co $\rightarrow$ $^{56}$Fe (0.0098 mag day$^{-1}$) decay rates indicating efficient gamma ray and/or positron trapping. For a comparison we also list the very late time (230 to 380 days) decay rates of SN 2005hk in Table \ref{tab:decay rate}. The behaviour of late time light curve is an important tool to understand the magnetic field structure \citep{1980ApJ...237L..81C,1998ApJ...500..360R, 1999ApJS..124..503M,2001ApJ...559.1019M}. The configuration of magnetic field is associated with the fraction of deposited positron energy. The flattening seen in the late time light curve indicates trapping of significant fraction of positrons implying a strong and tangled magnetic field instead of weak magnetic field configuration \citep{2007A&A...470L...1S}.

We estimate the peak absolute magnitude of SN 2014dt to be $M$$_V$ = -18.33$\pm$0.02 mag. Based on the estimate of peak brightness, SN 2014dt is similar to SNe 2005hk \citep{2007PASP..119..360P} ($M$$_V$ = -18.08$\pm$0.29 mag) and SN 2012Z \citep{2015A&A...573A...2S} ($M$$_V$ = -18.50$\pm$0.09 mag). For SN 2014dt, \cite{2016MNRAS.461..433F} estimate  the peak absolute magnitude $M$$_V$ = -17.4$\pm$0.5 mag. This difference in peak absolute magnitude arises because of the different distances adopted by \cite{2015ApJ...798L..37F} and this work. 

\subsection{Evolution of colours}

\begin{figure}
	\begin{center}
		\includegraphics[scale=0.42]{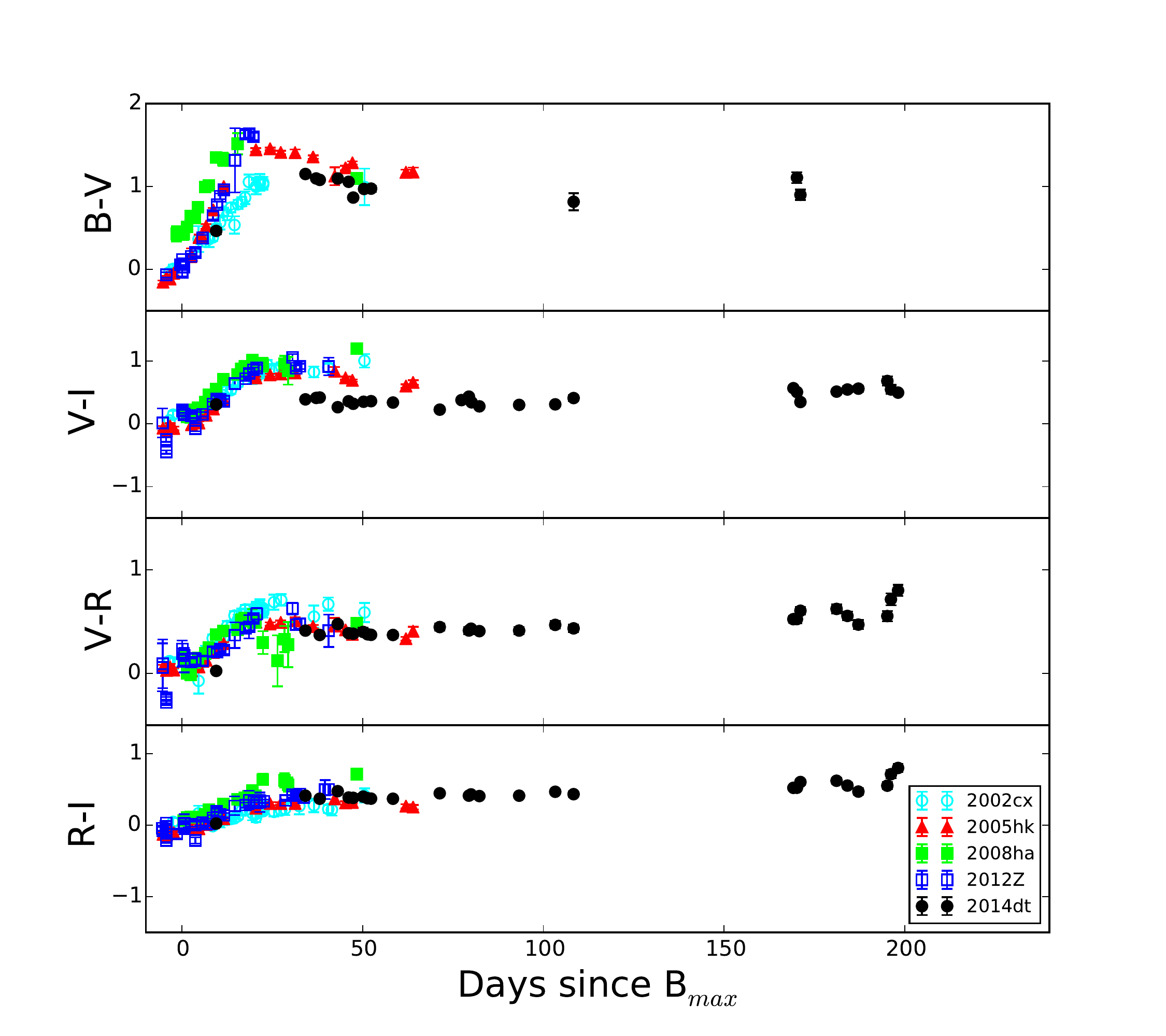}
	\end{center}
	\caption{The comparison of {\it B-V, V-I, V-R} and {\it R-I} colour evolution of SN 2014dt with four other Type Iax SNe. The error bars are smaller than the point size.}
	\label{fig:color_curve}
\end{figure}

The {\it B-V}, {\it V-I}, {\it V-R} and {\it R-I} colour curves of SN 2014dt are shown in Figure \ref{fig:color_curve} and compared with four other Type Iax SNe (see Table \ref{tab:photometric_parameters_different_SNe}). The colour curves of all SNe are corrected for reddening values given in Table \ref{tab:photometric_parameters_different_SNe}. From Figure \ref{fig:color_curve} we see that for most Type Iax SNe, the colour evolution is studied upto a time span of less than 100 days whereas in the case of SN 2014dt we have estimate of colours from 30 to 200 days. Between 30 to 70 day the {\it B-V} colours of SN 2014dt closely resembles the colour evolution of SN 2005hk with slightly bluer colours. {\it B-V} colour evolution of SN 2014dt follows a bluer trend upto $\sim$170 days. {\it V-I, V-R} and $R-I$ evolution are nearly constant upto $\sim$100 days and are redder at later phases. At phase $\sim$70 day {\it V-I, V-R} and $R-I$ colour evolution are similar to SN 2005hk. We notice some dip like features in $V-R$ and $R-I$ colours close to +200 days possibly resulting from a scatter in the $V,R$ and $I$ band data.

\subsection{Bolometric light curve}

\begin{figure}
	\begin{center}
		\includegraphics[scale=0.40]{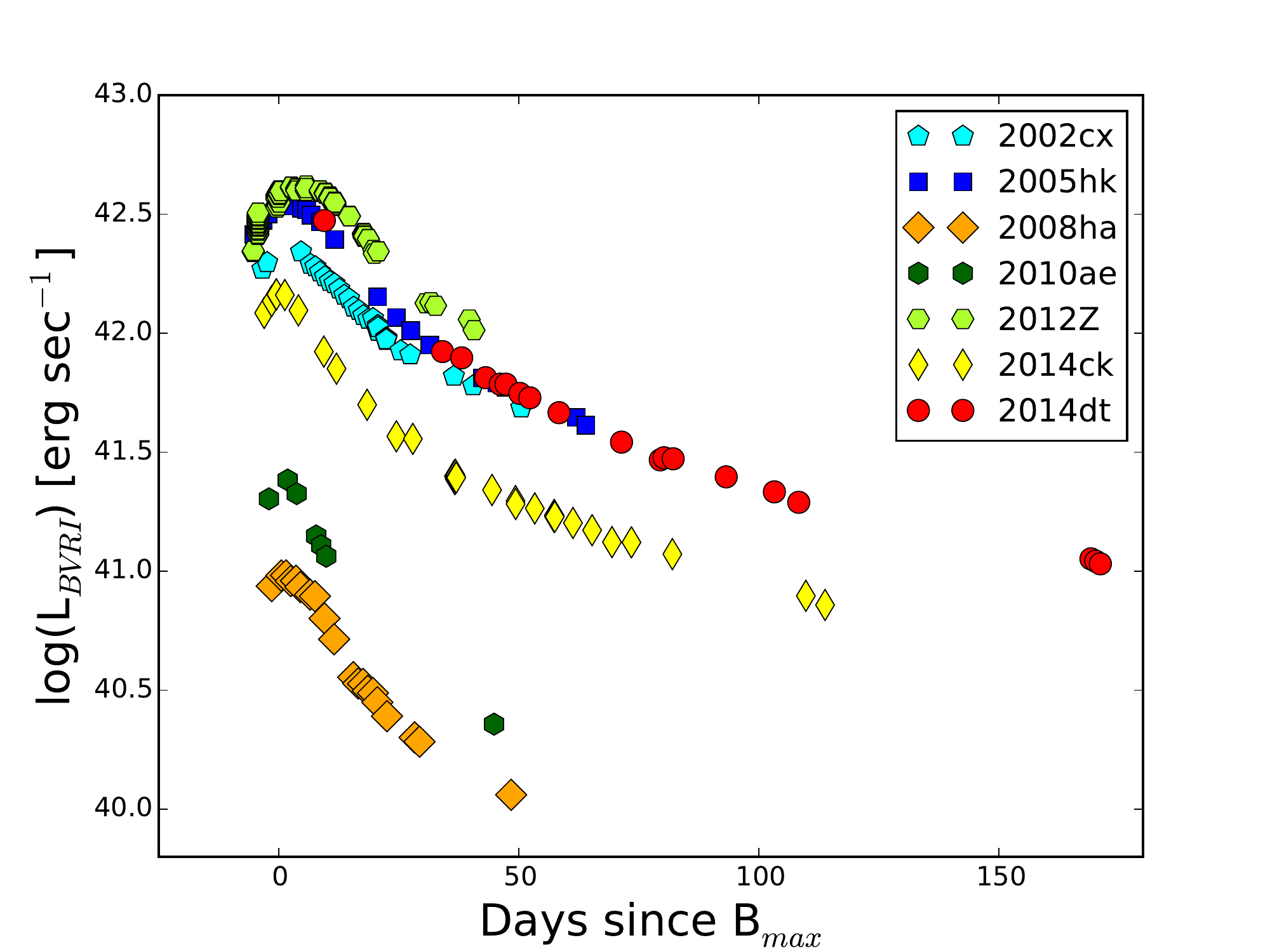}
	\end{center}
	\caption{$BVRI$ quasi bolometric light curve of SN 2014dt. A comparison with $BVRI$ quasi bolometric light curves of other Type Iax SNe (SNe 2002cx \citep{2003PASP..115..453L}, 2005hk \citep{2008ApJ...680..580S}, 2008ha \citep{2009AJ....138..376F}, 2010ae \citep{2014A&A...561A.146S}, 2012Z \citep{2015A&A...573A...2S,2015ApJ...806..191Y} and 2014ck \citep{2016MNRAS.459.1018T}) shows that the luminosity of SN 2014dt is similar to SN 2005hk.}.
	\label{fig:final_bolometric_plot}
\end{figure}

The light curves of Type Ia SNe are powered by the radioactive decay of $^{56}$Ni$\rightarrow$$^{56}$Co$\rightarrow$$^{56}$Fe
and the peak luminosity is directly related to amount of $^{56}$Ni synthesized in the explosion \citep{1982ApJ...253..785A}. Arnett's formulation is applied to the bolometric light curve for estimating various physical parameters like $^{56}$Ni mass, ejected mass {\it (M$_{ej}$)} and kinetic energy {\it (E$_k$)} of the explosion. The underlying assumptions of Arnett's model are homologous expansion of ejecta, dominance of radiation pressure, constant opacity, distribution of $^{56}$Ni peaking towards the ejected mass centre, small initial radius of explosion etc. \citep{1982ApJ...253..785A}. Most of the flux at early times emerges in the optical wavelength for a Type Ia SN and hence the quasi-bolometric luminosity obtained using optical bands gives a good estimate of the total bolometric luminosity {\it (UVOIR)}. We construct the quasi bolometric light curve of SN 2014dt by converting the extinction corrected {\it BVRI} magnitudes to fluxes by using zero points from \cite{1998A&A...333..231B}. The total flux in {\it BVRI} bands is estimated by integrating the flux between
{\it B} and {\it I} bands using trapezoidal rule and converted to luminosity by using the luminosity distance (assuming $\it{H}$$_o$ = 73$\pm$5 
km s$^{-1}$ Mpc$^{-1}$, $\Omega$$_m$ = 0.27,  $\Omega$$_v$ = 0.73). The quasi bolometric light curves of the SNe in our sample (SNe 2002cx \citep{2003PASP..115..453L}, 2005hk \citep{2008ApJ...680..580S}, 2008ha \citep{2009AJ....138..376F}, 2010ae \citep{2014A&A...561A.146S}, 2012Z \citep{2015A&A...573A...2S,2015ApJ...806..191Y} and 2014ck \citep{2016MNRAS.459.1018T}) are constructed in the same manner as discussed above for SN 2014dt. The quasi bolometric light curve of SN 2014dt is shown in Figure \ref{fig:final_bolometric_plot} along with a sample of Type Iax SNe. From the figure it is clear that the luminosity of SN 2014dt is comparable to SN 2005hk.

However the true bolometric luminosity results from the contribution of {\it UV}, optical and {\it IR} bands. In the case of SN 2014dt we only have the optical bands and hence we are missing contribution from the {\it UV} and {\it IR} bands. Since these peculiar events are rare it is difficult to apply a well defined correction for the missing bands which can be easily done in the case of Type Ia SNe  \citep{sent:1996,2000A&A...359..876C}. Both SNe 2005hk and 2012Z have $U$ and IR observations which can be used to estimate the contribution of $U$ and IR bands to the optical ({\it BVRI}) luminosity. We find that the total contribution of $U$ and {\it IR} bands is between 20-30\% at peak. \cite{2016MNRAS.459.1018T} also find that {\it U} and {\it IR} bands contribute  $\sim$ 35\% at peak in SN 2014ck. We apply Arnett's model to the quasi bolometric light curve of SN 2005hk (assuming constant optical opacity $k$$_{opt}$ = 0.1 cm$^2$/g and constant of integration $\beta$ = 13.6) and estimate the $^{56}$Ni mass to be 0.14 M$_{\odot}$, ejecta mass {\it M$_{ej}$} $=$ 0.98 M$_{\odot}$ and kinetic energy {\it E$_{k}$} $=$ 0.41 $\times$ 10$^{51}$ erg for photospheric velocity of 6500 km sec$^{-1}$. After correcting for the missing $U$ and {\it IR} bands flux (as discussed above) of SN 2005hk, we estimate the $^{56}$Ni mass to be 0.19 M$_{\odot}$. The results of Arnett's model are in good agreement with those found in \cite{2008ApJ...680..580S}, \cite{2007PASP..119..360P} and \cite{2015A&A...573A...2S} for SN 2005hk.  The ejecta masses of SN 2005hk that are estimated by \cite{2008ApJ...680..580S} (Ch-mass ejecta, photospheric velocity 6500 to 6000 km sec$^{-1}$ at -6 and +3 day respectively), \cite{2015A&A...573A...2S} (ejecta mass between 1.5 - 2.0 M$_{\odot}$, photospheric velocity between 5000 to 7000 km sec$^{-1}$) and this work vary because of the values of the photospheric velocity used in the calculation and the different methods used. We have taken the upper limit  Keeping in view the remarkable similarities between SN 2005hk and SN 2014dt we can say that a comparable amount of $^{56}$Ni would have been ejected in the explosion of SN 2014dt. Using the analytical expressions in \cite{2003ApJ...593..931M} and \cite{2015ApJ...806..191Y}, we fit a late phase energy deposition function to the nebular phase bolometric light curve of SN 2014dt and estimate the ejecta mass to be 0.95 M$_{\odot}$  which is in agreement with the values obtained using Arnett's model at early times.

%%%%%%%%%%%%%%%%%%%%%%%%%%%%%%%%%%%%%%

\section{Spectral Evolution}

\begin{figure}
	\begin{center}
		\includegraphics[scale=0.24]{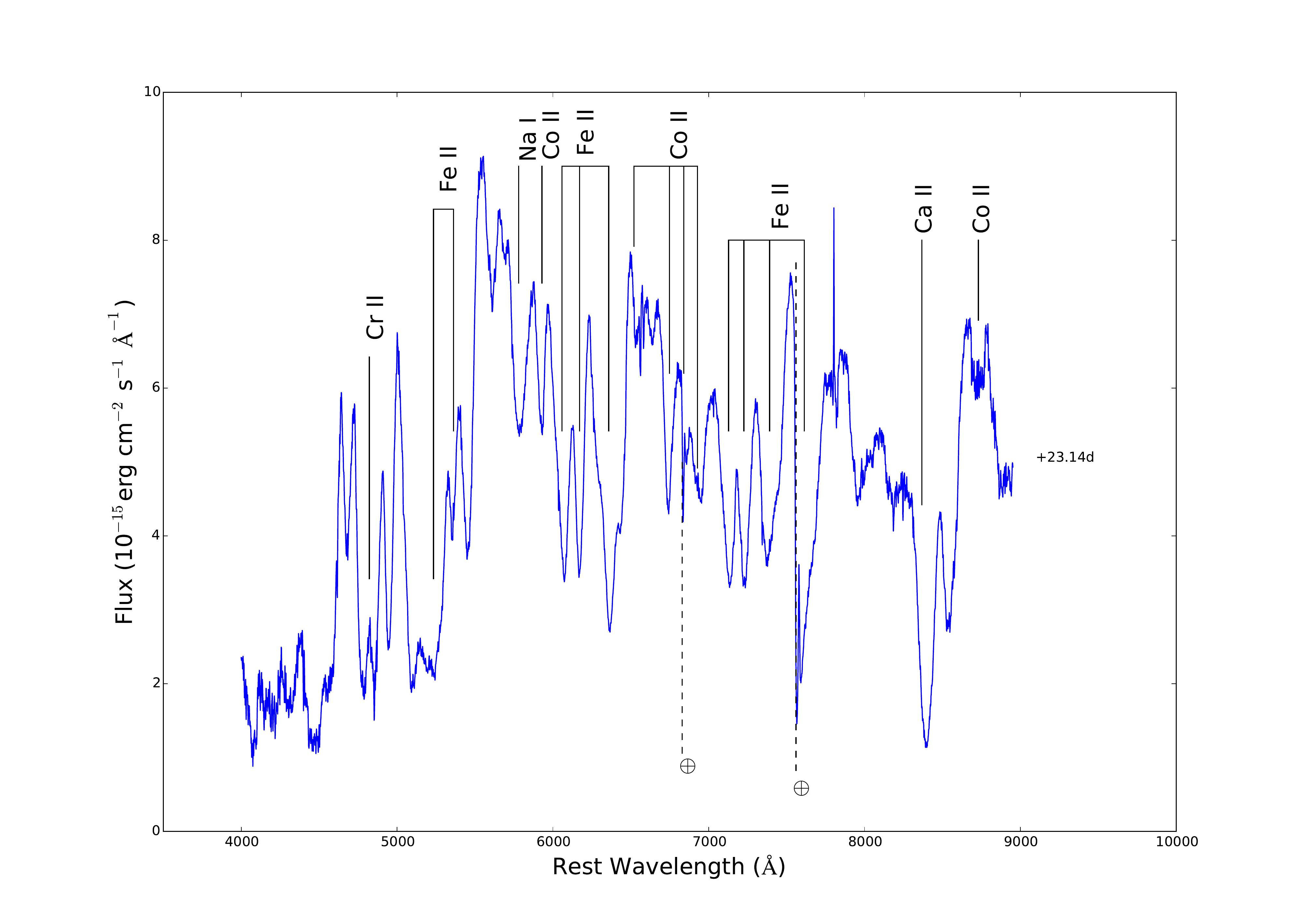}
	\end{center}
	\caption{The earliest spectrum of SN 2014dt at 23 days is shown in the figure. Phase is calculated with respect to {\it B}$_{max}$. Line identification is done using \citet{2004PASP..116..903B}. Telluric features are also shown in the Figure.} 
	\label{fig:23day_spectrum}
\end{figure} 

\begin{figure*}
	\begin{center}
		\includegraphics[scale=0.50]{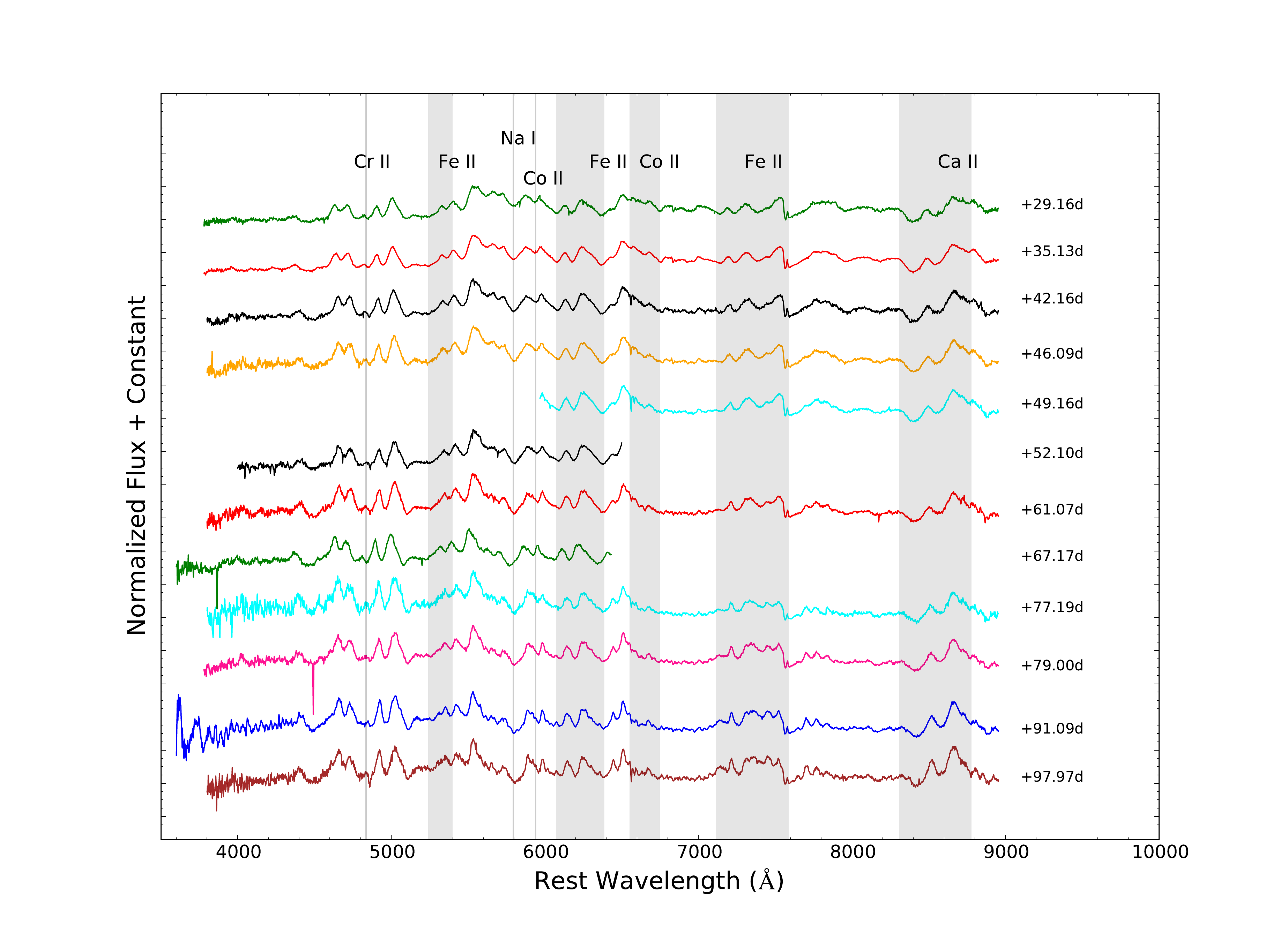}
	\end{center}
	\caption{Spectral evolution of SN 2014dt during 29 days to 98 days relative to {\it B}$_{max}$. Line identification is based on \citealt{2004PASP..116..903B}. The different lines are indicated by shaded regions.}
	\label{fig:spectra_second_plot}
\end{figure*}

\begin{figure*} 
	\begin{center}
		%\resizebox{20.0cm}{10.0cm}{\includegraphics{spectra2014dt1.pdf}}
		\includegraphics[scale=0.50]{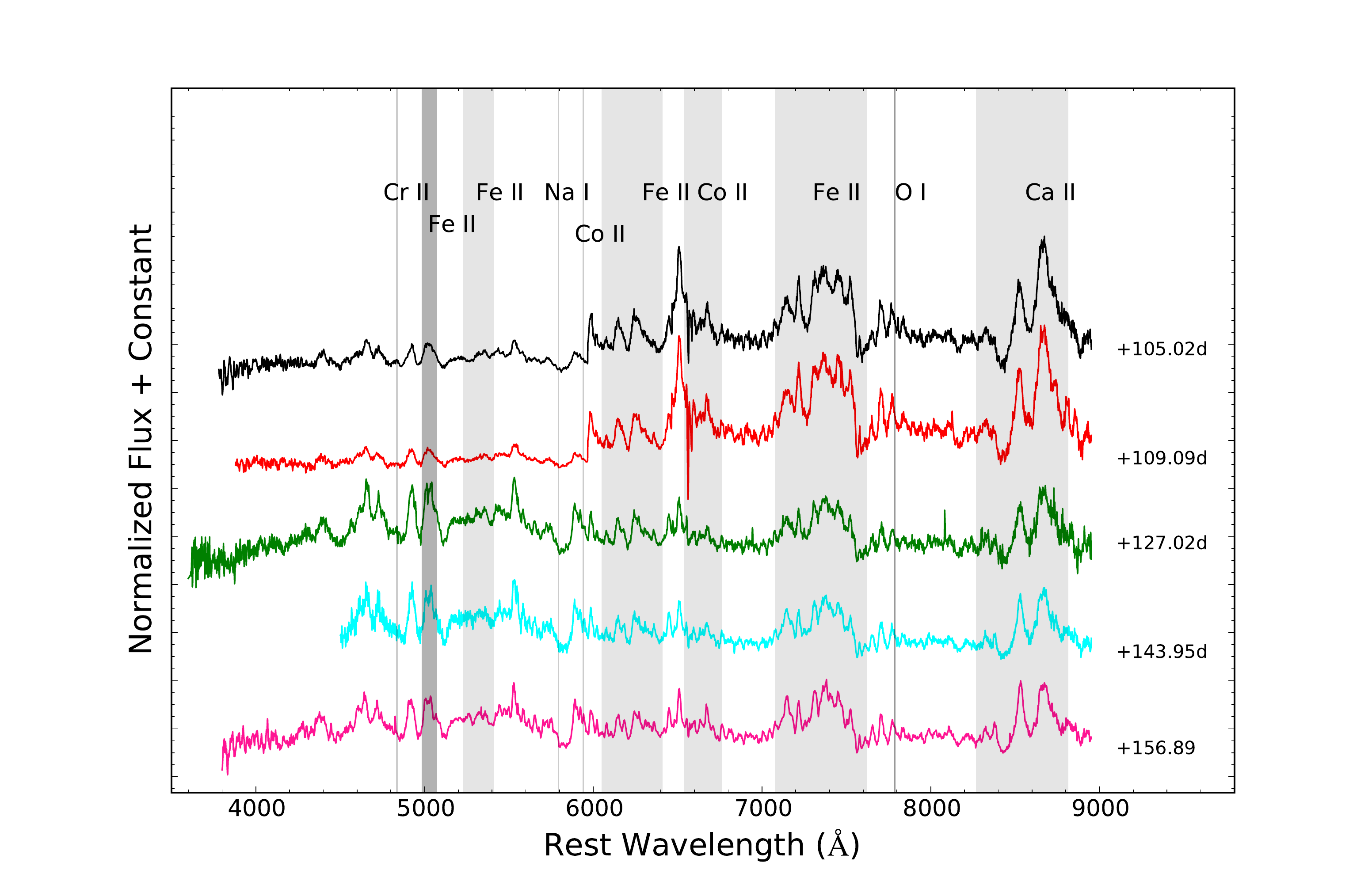}
	\end{center}
	\caption{Spectral evolution of SN 2014dt from phase 105 days to 157 days relative to $B$$_{max}$. Line identification is based on \citet{2004PASP..116..903B}. The different lines are indicated by shaded regions.}
	\label{fig:spectra_2014dt3}
\end{figure*}

We present spectral evolution at 18 epochs between 23 to 157 days after {\it B}$_{max}$ in Figures \ref{fig:23day_spectrum}, \ref{fig:spectra_second_plot} and \ref{fig:spectra_2014dt3}. Telluric features have not been removed from the spectra and are marked in Figure \ref{fig:23day_spectrum}. Line identification has been done following \cite{2004PASP..116..903B}. We run SNID \citep{2007ApJ...666.1024B} spectral identification code on our early phase spectrum which corresponds to 23 days since $B$$_{max}$. The earliest spectrum at 23 days (Figure \ref{fig:23day_spectrum}) is dominated by Fe II and Co II lines. Features due to Cr II ($\sim$4800 \AA) and Na ID (5890 \AA) line are also present. In the red region NIR features due to Ca II (NIR triplet) and Co II are also seen. The Si II line at $\sim$6150 \AA~ is not seen in our first spectrum obtained at 23 days. This line is the identifying feature of Type Ia SNe during early photospheric phase which gradually disappears and gets blended with other lines like Fe, Co etc. In Figure \ref{fig:spectra_second_plot} spectral evolution during 29 to 98 days is presented. Around 42 days lines due to Co II in the wavelength range 6500 \AA~to 6800 \AA~ evolve fast in comparison to other lines. The spectral evolution between 105 to 157 days shows an increase in the emission lines (Figure \ref{fig:spectra_2014dt3}) as is expected in a SN spectra when the ejecta becomes optically thin. The late phases mark the disappearance of Cr II lines around 4800 \AA~ however new features due to Fe II multiplets ($\lambda$ 5000 \AA) and O I ($\lambda$ 7773 \AA) appear in the spectra. The blue continuum weakens as the spectra evolves.

Type Iax SNe form a distinct class and each SN shows some unique features. We have compared spectral evolution at different phases with other well studied SNe 2002cx \citep{2003PASP..115..453L}, 2005hk \citep{2008ApJ...680..580S}, {2008ge \citep{2010AJ....140.1321F}}, 2012Z \citep{2015A&A...573A...2S} and 2015H \citep{2016A&A...589A..89M}. 

\begin{figure*}
	\begin{center}
		\includegraphics[width=15cm,height=10cm]{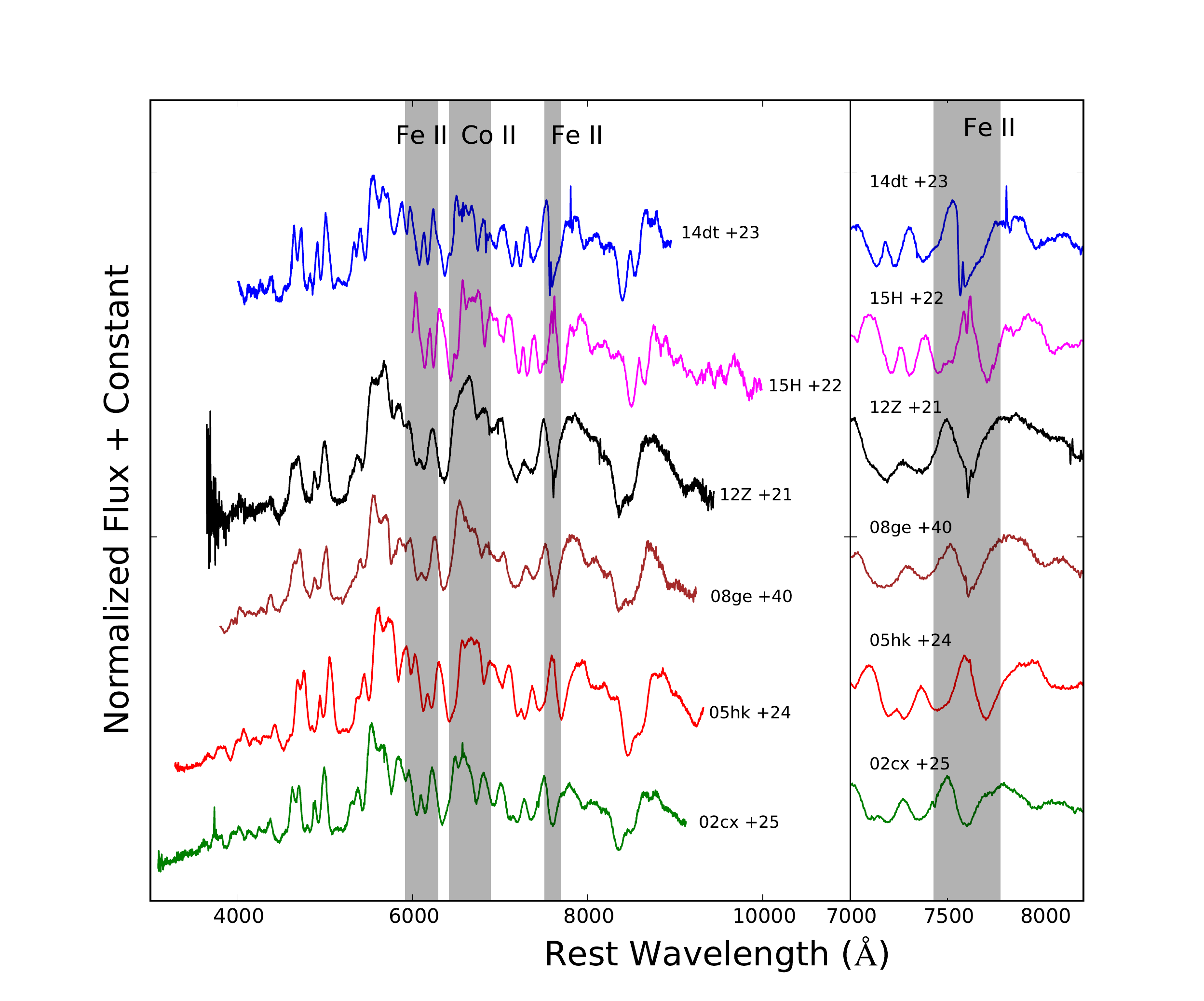} 
		\includegraphics[width=15cm,height=10cm]{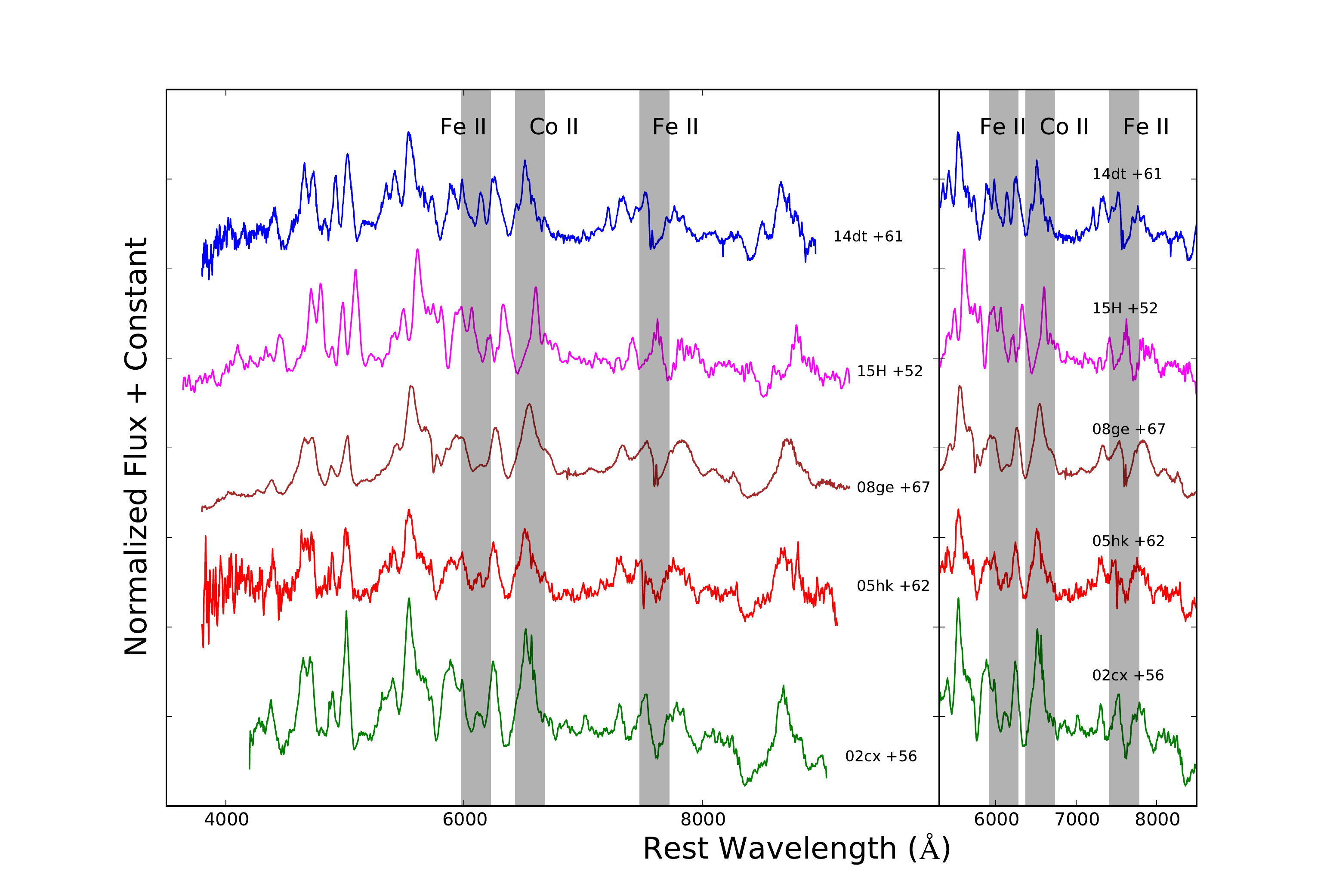} 
		
	\end{center}
	\caption{A comparison of SN 2014dt spectra at different epochs with other well studied Type Iax SNe. All spectra have been shifted on a relative scale and shifted vertically for clarity. {\bf Top figure:} A comparison of SN 2014dt at phase 23 days is shown. {\bf Bottom figure:} At 61 days there is overall a similarity exists between all the SNe except some line forming regions shown by dark shade. Also demarcating features are presented in a zoomed view in Figure. A full description of the observed features has been given under section 6.}
	\label{fig:inset_plot}
\end{figure*}

The earliest spectrum (23 days) of SN 2014dt has a good match at different line forming regions with all the five SNe used in comparison (Figure \ref{fig:inset_plot} (top figure)). The two absorption dips due to Fe II lines at 7308 \AA~ and 7462 \AA~ are clearly seen in SNe 2012Z and 2014dt, a slight hint in SN 2015H and single absorption dip in SNe 2005hk and 2002cx (presented in zoomed view in Figure \ref{fig:inset_plot}). Since we have not done telluric correction, this may be the region where telluric line 7603 \AA~also comes. In the case of SN 2014dt it is clearly affected by telluric lines. The Co triplet at wavelength range 6500 \AA~ to 6800 \AA~ in SN 2014dt best matches with SN 2015H in comparison to other SNe (Figure \ref{fig:inset_plot}).

At 61 days the overall spectrum of SN 2014dt is similar to other Type Iax SNe. Fe II multiplet near 6000 \AA~of SN 2014dt matches well with other SNe 2002cx, 2005hk and SN 2015H, only SN 2008ge has a poor resemblance with SN 2014dt in this wavelength regime. Co II features at 6521 \AA~ and 6578 \AA~ are clearly visible in the case of SN 2002cx, two closely spaced peak are seen in the case of SN 2005hk, single absorption peak in the case of SNe 2008ge and 2015H, whereas in SN 2014dt the emission lines are marginally double peaked. The comparison plot along with the zoomed view of differences is shown in Figure \ref{fig:inset_plot} (bottom figure).

\begin{figure} 
	\begin{center}
		\includegraphics[scale=0.30]{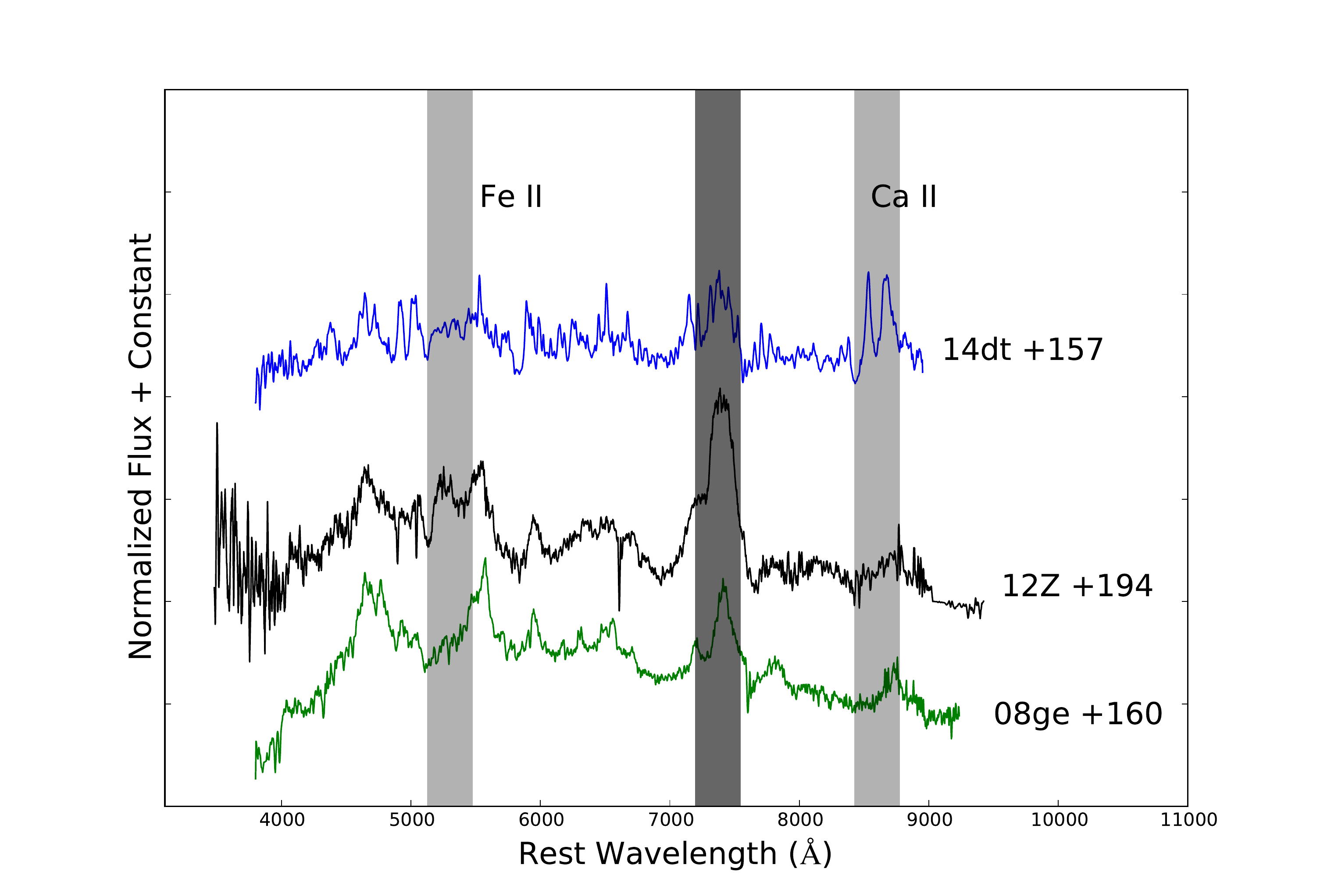}
	\end{center}
	\caption{Comparison of late nebular phase spectrum of SN 2014dt with SNe 2008ge and 2012Z. The most noticeable features around 7300 \AA~and 7400 \AA~are shown by a relatively dark shade. In the case of SNe 2008ge and 2012Z there are broad emission features with blending of a considerable number of ions. In the case of SN 2014dt there is no prominent blending so line identification is easy. Fe II multiplet at 5200 \AA~and 5400 \AA~along with NIR triplet are present in all three SNe 2014dt, 2012Z and 2008ge.}
	\label{fig:157day_plot}
\end{figure}

Late-time spectrum (157 day) of SN 2014dt is compared with SNe 2008ge and 2012Z in Figure \ref{fig:157day_plot}. The overall spectral features of SN 2008ge are similar to SN 2014dt except that the feature at $\sim$ 7300 \AA~ (identified as [FeII]) which increases in strength with time \citep{2010AJ....140.1321F}. This broad emission feature is a result of blending of [Fe II] line with other lines. In SN 2014dt the blending is less than seen in SNe 2008ge and 2012Z. The region of interest, discussed above is marked with a dark shaded region in Figure \ref{fig:157day_plot}. The NIR triplet is also present in SNe 2014dt, 2008ge and 2012Z along with different blending factors as discussed above. All the lines in SN 2014dt are narrower than SNe 2008ge and 2012Z.   
 
\subsection{Velocity Evolution}

\begin{figure}
	\begin{center}
		\includegraphics[scale=0.45]{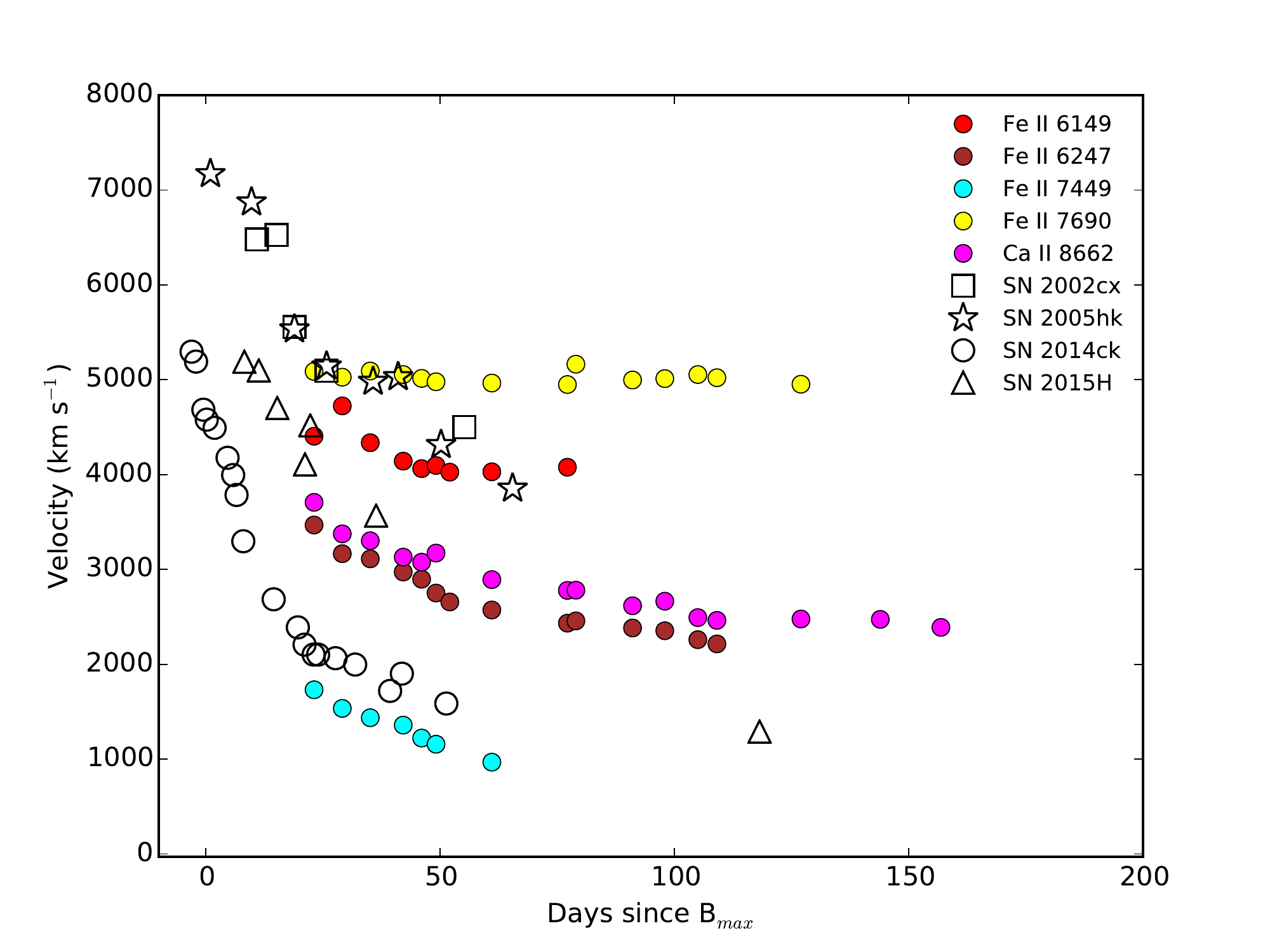}
	\end{center}
	\caption{Absorption minima evolution of a few lines for SN 2014dt and its comparison with SNe 2002cx, 2005hk, 2014ck and 2015H.}
	\label{fig:velocity_plot}
\end{figure}

We estimate the velocities of a few lines Fe II 6149 \AA, Fe II 6247 \AA, Fe II 7449 \AA, Fe II 7690 \AA~and Ca II 8662 \AA~by measuring the absorption minima (Figure \ref{fig:velocity_plot}). Spectral evolution of SN 2014dt is enriched with Fe II lines and their evolution is significantly prominent. All the estimated line velocities are nearly half of the velocities usually found in typical Type Ia SNe. In the case of SN 2014dt Fe II 7449 \AA~ line shows a linear decline (velocity falls from 1700 to 1100 km sec$^{-1}$) upto $\sim$50 days and other lines like Fe II 6149 \AA, Fe II 6247 \AA, Fe II 7690 \AA~and Ca II 8662 \AA~show a linear decline  followed by a nearly constant evolution. Velocity gradient for all the lines are very small during the entire spectral sequence. In Figure \ref{fig:velocity_plot}, we also show the velocity evolution measured from Fe II 6149 \AA~for SNe 2002cx, 2005hk, 2015H \citep{2016A&A...589A..89M} and 2014ck \citep{2016MNRAS.459.1018T}. At the phase of SN 2014dt beyond 20 days, SNe 2002cx and 2005hk both show higher velocity than SN 2014dt whereas SNe 2014ck and 2015H both are associated with lower velocities as compared to SN 2014dt. Velocities associated with SNe 2002cx and 2005hk are nearly 500 km sec$^{-1}$ higher than velocities associated with SN 2014dt. SNe 2014ck and 2015H are associated with lower velocities nearly 2500 km sec$^{-1}$ and 500 km sec$^{-1}$ respectively than velocities associated with SN 2014dt. The estimated velocities of different lines is indicative of the fact that explosion energy associated with SN 2014dt is low (0.24 $\times$ 10$^{51}$ erg for a photospheric velocity of 5000 km sec$^{-1}$ and ejecta mass 0.98 M$_{\odot}$) as compared to normal Type Ia SNe (1.39 $\times$ 10$^{51}$ erg, \cite{2016ApJ...832...13W}). If we consider a M$_{ch}$ white dwarf then for a photospheric velocity of 5000 km sec$^{-1}$ the kinetic energy is 0.35 $\times$ 10$^{51}$ erg which is higher than that obtained for SN 2014dt.

\subsection {Spectral Modeling} 

\begin{figure*}
	\begin{center}
		\includegraphics[scale=0.50]{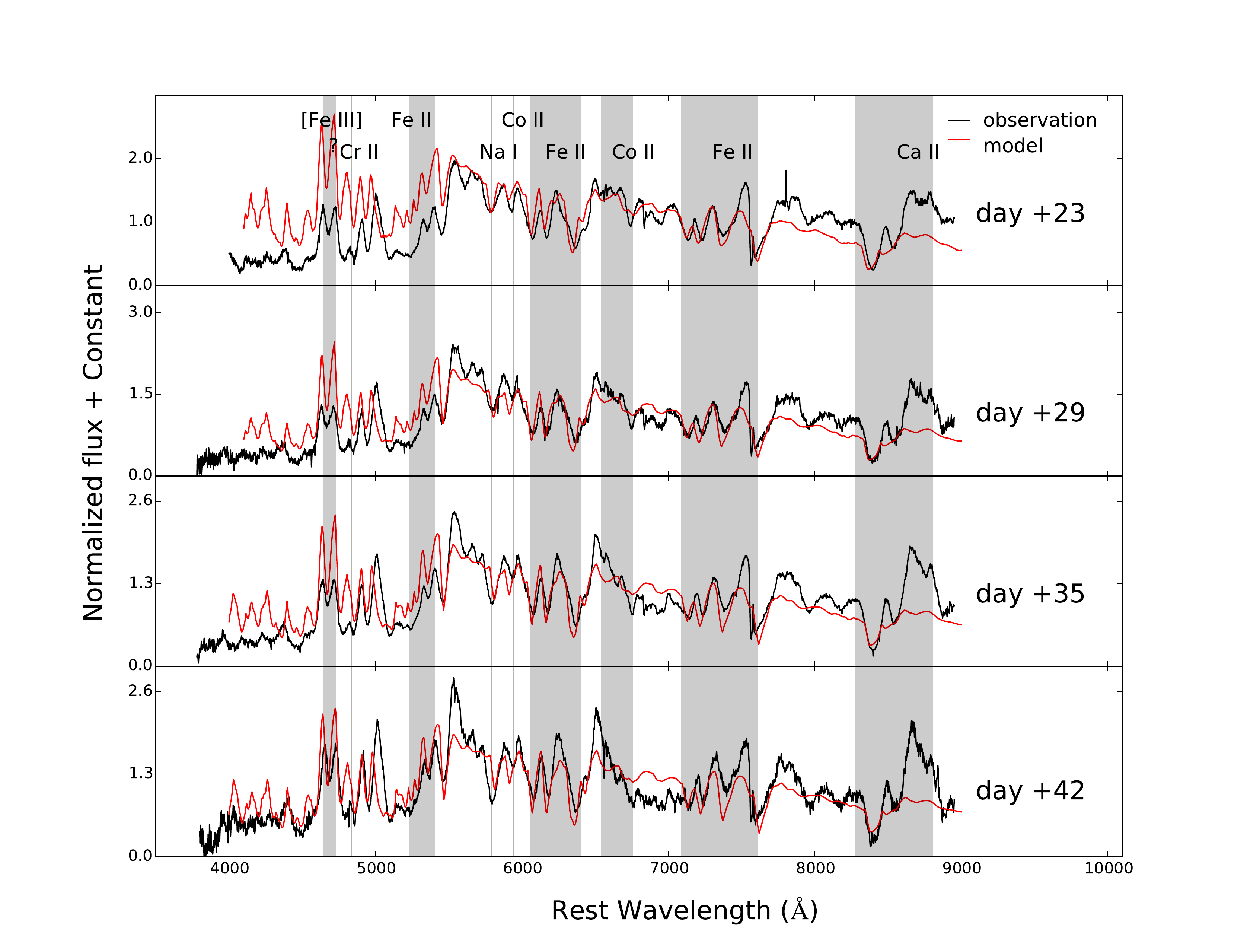}
	\end{center}
	\caption{A comparison between early phase observed spectra of SN 2014dt and synthetic spectra generated by SYN++ code. Prominent absorption features are marked with shaded region.}
	\label{fig:spectral_modelling}
\end{figure*}

We have used parameterized spectrum synthesis code SYN++  \citep{2007AIPC..924..342B, 2011PASP..123..237T} for spectral modeling. We model the first four epochs (23, 29, 35 and 42 days since $B$$_{max}$) of SN 2014dt with spectral synthesis code SYN++. In figure \ref{fig:spectral_modelling}, we present the observed spectra with the output synthetic spectra generated by the SYN++ code. It was not possible to fit some of the broad emission features due to LTE approximation and resonance scattering of this code. Fe II and Co II features match very well. The Fe [III] doublet at 4700 \AA~Co II and Fe II lines are fitted well. Co II and Fe II lines between 6700 \AA~and 7700 \AA~are also well produced. Although the NIR region is difficult to reproduce, the dip due to Ca is visible in the modelled spectra. As the phase increases, absorption profiles become sharper and deeper with the photospheric velocities gradually decreasing from 4500 km sec$^{-1}$ to 3800 km sec$^{-1}$. For fitting all the later phases, same species and parameters (as used in the 23 day spectrum) with gradual change are used which indicates slow evolution of ejecta configurations. Boltzmann excitation temperature varies between 7000 K to 4400 K for different lines. Also, the range for blackbody temperature was 8000 K to 6600 K.   

\section{Summary}

SN 2014dt belongs to the subclass of Type Iax SNe and resembles other Type Iax SNe. Since SN 2014dt was discovered post peak, template fitting method yields a best match to the light curves of SN 2005hk with an estimated peak magnitude of 13.59$\pm$0.04 mag in $B$ band and epoch of maximum JD = 2456950.34. Late phase light curves of SN 2014dt show some flattening which is explained by gamma ray and/or positron trapping. $B-V$ colour of SN 2014dt is marginally bluer than SN 2005hk, however the overall colour evolution of SN 2014dt is similar to SN 2005hk. 

\begin{figure*}
	\begin{center}
		\includegraphics[scale=0.90]{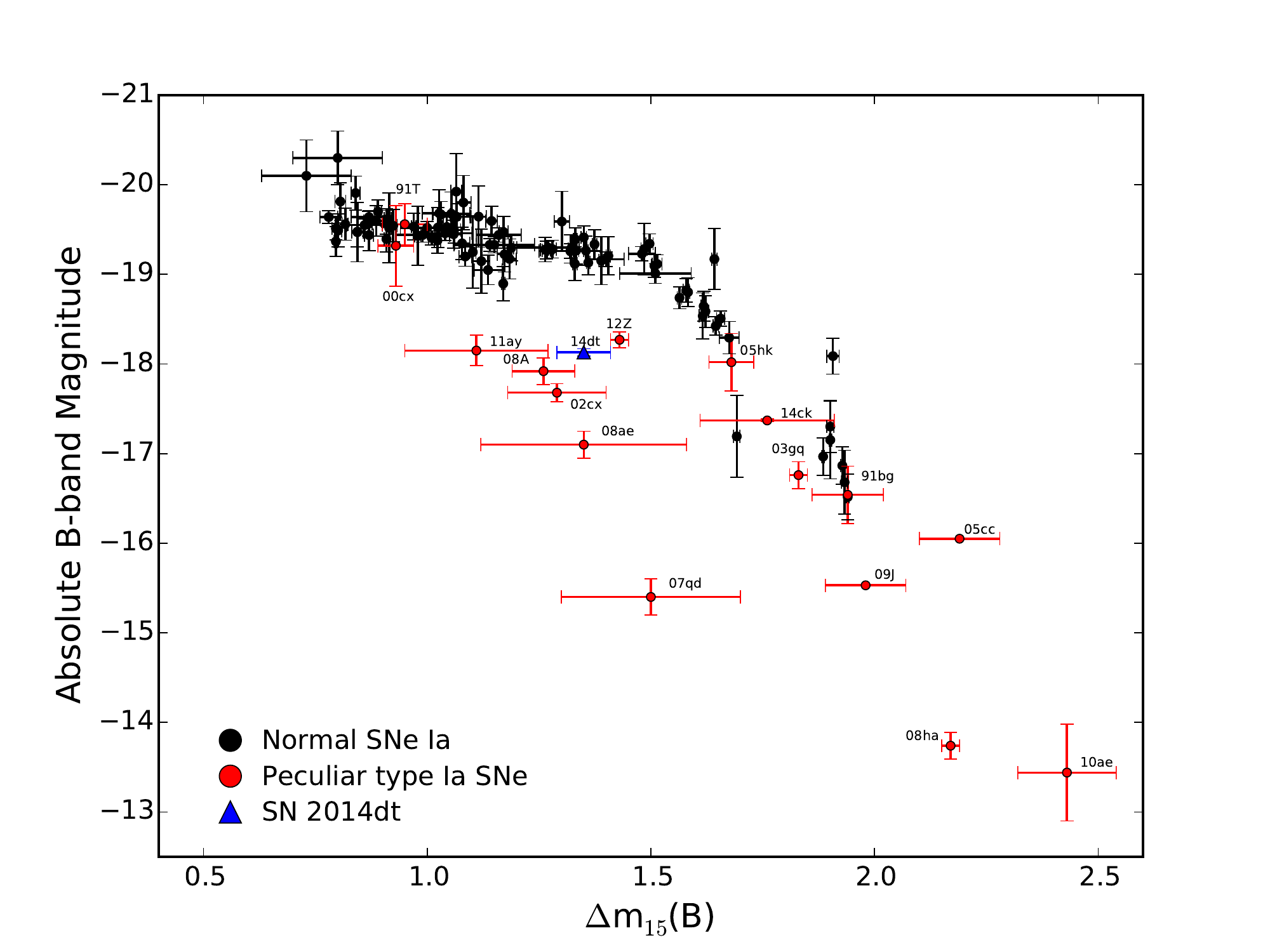}
	\end{center}
	\caption{Absolute $B$ band magnitude versus decline rate $\Delta m_{15}$ for a sample of Type Iax SNe 1991T \citep{2001ApJ...547L.103G}, 1991bg \citep{1996MNRAS.283....1T},  2000cx \citep{2001PASP..113.1178L}, 2002cx \citep{2003PASP..115..453L}, 2003gq \citep{2013ApJ...767...57F}, 2005cc (\citep{2013ApJ...767...57F}, our computed value), 2005hk \citep{2008ApJ...680..580S}, 2007qd \citep{2010ApJ...720..704M}, 2008A \citep{2013ApJ...767...57F}, 2008ae \citep{2013ApJ...767...57F}, 2008ha \citep{2009AJ....138..376F}, 2010ae \citep{2014A&A...561A.146S}, 2011ay \citep{2015MNRAS.453.2103S}, 2012Z \citep{2015A&A...573A...2S}, 2014ck \citep{2016MNRAS.459.1018T} and 2014dt (this work). Normal Type Ia SNe \citep{2006AJ....131..527J,2007ApJ...659..122J,2010ApJS..190..418G,2015A&A...573A...2S} are also shown in the Figure.}
	\label{fig:delta_m15}
\end{figure*}

On the basis of the estimated peak absolute magnitude ($M$$_V$ = -18.33$\pm$0.02 mag), we can conclusively say that SN 2014dt falls in the category of bright Type Iax SNe like SN 2005hk and SN 2012Z. The same can be inferred from Figure \ref{fig:delta_m15} which is a scatter plot of absolute magnitude in $B$ band ($M$$_B$) and $\Delta$m$_{15}$(B) of a number of Type Iax SNe and a sample of Type Ia SNe. The photometric properties of SN 2014dt are strikingly similar to those of SN 2005hk hence we can say that the explosion parameters of SN 2014dt will be comparable to those of SN 2005hk. For SN 2005hk, we estimate $^{56}$Ni between 0.14--0.19 M$_\odot$ using Arnett's method with ejecta mass {\it M$_{ej}$} $=$ 0.98 M$_{\odot}$ and kinetic energy of explosion {\it E$_{k}$} $=$ 0.41 $\times$ 10$^{51}$ erg. We suggest that the amount of synthesized $^{56}$Ni in the explosion of SN 2014dt would be at least 0.14 M$_\odot$. Using a late phase energy deposition function in the nebular phase bolometric light curve of SN 2014dt we find the ejecta mass to be 0.95 M$_\odot$. Low explosion energy and low $^{56}$Ni mass associated with SN 2014dt makes it less luminous as compared to Type Ia SNe.

Our spectroscopic campaign started 23 days since $B$$_{max}$. Spectroscopic features of SN 2014dt have fairly good resemblance with other Type Iax SNe. The observed spectra are modelled with lower black body temperature (8000-6600 K) and lower photospheric velocities (4500 km sec$^{-1}$ to 3800 km sec$^{-1}$). Velocity evolution feature also confirms low energy budget of SN 2014dt, a typical feature of peculiar Type Iax SNe. Velocity of absorption minima for a number of lines vary between $\sim$ 1500 km sec$^{-1}$ to $\sim$ 5500 km sec$^{-1}$.

 SN 2014dt has pre-explosion {\it HST} images and second deepest after SN 2012Z. With pre-explosion images of SN 2014dt, \cite{2015ApJ...798L..37F} did not detect any stellar source at the SN location to relatively deep limits and conclude that progenitor system of SN 2014dt was less luminous in $B$ band than SN 2012Z. The plausible progenitor of SN 2014dt could be a carbon oxygen white dwarf primary/helium-star companion, similar to SN 2012Z with a slightly smaller or hotter donor \citep{2015ApJ...798L..37F}. The IR excess seen in the late-time light curves implies the association of the progenitor with a pre-existing dust shell, which includes a binary system with a red giant, a red supergiant or an asymptotic giant branch star as a mass donor \citep{2016ApJ...816L..13F}. \cite{2016MNRAS.461..433F} claimed on the basis of their existing data that the observed IR excess can not be accounted by dust emission rather the plausible reason for strong IR flux seen about 315 days after maximum brightness is from a bound remnant with an extended optically thick super-Eddington wind. The bound remnant mechanism is consistent with the late-time data of Type Iax SNe and could also be valid for SN 2014dt.

Continued late-time observations of SN 2014dt will be necessary for putting constraints on the progenitor system and to conclude if the emission is dominated by the bound remnant. In many cases, the chances of late-time detection are hampered by object's distance, its faintness or its location close to a bright host galaxy nucleus. The proximity of SN 2014dt and its location in the host galaxy is ideal for a late-time detection and makes this object an excellent candidate for long term monitoring.

\section*{Acknowledgments}

We would like to extend our gratitude to the anonymous referee for providing useful comments and suggestions to substantially improve the manuscript. BK acknowledges the Science and Engineering Research Board (SERB) under the Department of Science \& Technology, Govt. of India, for financial assistance in the form of National Post-Doctoral Fellowship (Ref. no. PDF/2016/001563). One of the authors MS thanks and acknowledges Ms. Arti Joshi and Ms. Sapna Mishra for their valuable support and discussion. The authors are grateful to Lina Tomasella for providing the data files of bolometric light curves of different SNe. We thank M. Stritzinger for providing $\Delta$m$_{15}$(B) and {\it B}$_{max}$ values. 
We thank the observing staff and observing assistants at 104 cm ST, 130 cm DFOT and 201cm HCT for their support during observations of SN 2014dt. We acknowledge Wiezmann Interactive Supernova data REPository http://wiserep.weizmann.ac.il (WISeREP) \citep{2012PASP..124..668Y}. This research has made use of the CfA Supernova Archive, which is funded in part by the National Science Foundation through grant AST 0907903. 

%%%%%%%%%%%%%%%%%%%%%%%%%%%%%%%%%%%%%%%%%%%%%%%%%%



%\bibliographystyle{mnras}

%\begin{thebibliography}{99}
%\end{thebibliography}

%%%%%%%%%%%%%%%%%%%%%%%%%%%%%%%%%%%%%%%%%%%%%%%%%%

%%%%%%%%%%%%%%%%% APPENDICES %%%%%%%%%%%%%%%%%%%%%

%\appendix



%%%%%%%%%%%%%%%%%%%%%%%%%%%%%%%%%%%%%%%%%%%%%%%%%%

\bibliographystyle{mnras}
\bibliography{refag}

\begin{thebibliography}{}
\makeatletter
\relax
\def\mn@urlcharsother{\let\do\@makeother \do\$\do\&\do\#\do\^\do\_\do\%\do\~}
\def\mn@doi{\begingroup\mn@urlcharsother \@ifnextchar [ {\mn@doi@}
  {\mn@doi@[]}}
\def\mn@doi@[#1]#2{\def\@tempa{#1}\ifx\@tempa\@empty \href
  {http://dx.doi.org/#2} {doi:#2}\else \href {http://dx.doi.org/#2} {#1}\fi
  \endgroup}
\def\mn@eprint#1#2{\mn@eprint@#1:#2::\@nil}
\def\mn@eprint@arXiv#1{\href {http://arxiv.org/abs/#1} {{\tt arXiv:#1}}}
\def\mn@eprint@dblp#1{\href {http://dblp.uni-trier.de/rec/bibtex/#1.xml}
  {dblp:#1}}
\def\mn@eprint@#1:#2:#3:#4\@nil{\def\@tempa {#1}\def\@tempb {#2}\def\@tempc
  {#3}\ifx \@tempc \@empty \let \@tempc \@tempb \let \@tempb \@tempa \fi \ifx
  \@tempb \@empty \def\@tempb {arXiv}\fi \@ifundefined
  {mn@eprint@\@tempb}{\@tempb:\@tempc}{\expandafter \expandafter \csname
  mn@eprint@\@tempb\endcsname \expandafter{\@tempc}}}

\bibitem[\protect\citeauthoryear{{Arnett}}{{Arnett}}{1982}]{1982ApJ...253..785A}
{Arnett} W.~D.,  1982, \mn@doi [\apj] {10.1086/159681}, \href
  {http://adsabs.harvard.edu/abs/1982ApJ...253..785A} {253, 785}

\bibitem[\protect\citeauthoryear{{Bessell}, {Castelli}  \& {Plez}}{{Bessell}
  et~al.}{1998}]{1998A&A...333..231B}
{Bessell} M.~S.,  {Castelli} F.,   {Plez} B.,  1998, \aap, \href
  {http://adsabs.harvard.edu/abs/1998A%26A...333..231B} {333, 231}

\bibitem[\protect\citeauthoryear{{Binggeli}, {Sandage}  \&
  {Tammann}}{{Binggeli} et~al.}{1985}]{1985AJ.....90.1681B}
{Binggeli} B.,  {Sandage} A.,   {Tammann} G.~A.,  1985, \mn@doi [\aj]
  {10.1086/113874}, \href {http://adsabs.harvard.edu/abs/1985AJ.....90.1681B}
  {90, 1681}

\bibitem[\protect\citeauthoryear{{Blondin} \& {Tonry}}{{Blondin} \&
  {Tonry}}{2007}]{2007ApJ...666.1024B}
{Blondin} S.,  {Tonry} J.~L.,  2007, \mn@doi [\apj] {10.1086/520494}, \href
  {http://adsabs.harvard.edu/abs/2007ApJ...666.1024B} {666, 1024}

\bibitem[\protect\citeauthoryear{{Bose} \& {Kumar}}{{Bose} \&
  {Kumar}}{2014}]{2014ApJ...782...98B}
{Bose} S.,  {Kumar} B.,  2014, \mn@doi [\apj] {10.1088/0004-637X/782/2/98},
  \href {http://adsabs.harvard.edu/abs/2014ApJ...782...98B} {782, 98}

\bibitem[\protect\citeauthoryear{{Bottinelli}, {Gouguenheim}, {Paturel}  \& {de
  Vaucouleurs}}{{Bottinelli} et~al.}{1984}]{1984A&AS...56..381B}
{Bottinelli} L.,  {Gouguenheim} L.,  {Paturel} G.,   {de Vaucouleurs} G.,
  1984, \aaps, \href {http://adsabs.harvard.edu/abs/1984A%26AS...56..381B} {56,
  381}

\bibitem[\protect\citeauthoryear{{Branch}, {Baron}, {Thomas}, {Kasen}, {Li}  \&
  {Filippenko}}{{Branch} et~al.}{2004}]{2004PASP..116..903B}
{Branch} D.,  {Baron} E.,  {Thomas} R.~C.,  {Kasen} D.,  {Li} W.,
  {Filippenko} A.~V.,  2004, \mn@doi [\pasp] {10.1086/425081}, \href
  {http://adsabs.harvard.edu/abs/2004PASP..116..903B} {116, 903}

\bibitem[\protect\citeauthoryear{{Branch}, {Parrent}, {Troxel}, {Casebeer},
  {Jeffery}, {Baron}, {Ketchum}  \& {Hall}}{{Branch}
  et~al.}{2007}]{2007AIPC..924..342B}
{Branch} D.,  {Parrent} J.,  {Troxel} M.~A.,  {Casebeer} D.,  {Jeffery} D.~J.,
  {Baron} E.,  {Ketchum} W.,   {Hall} N.,  2007, in {di Salvo} T.,  {Israel}
  G.~L.,  {Piersant} L.,  {Burderi} L.,  {Matt} G.,  {Tornambe} A.,   {Menna}
  M.~T.,  eds,  American Institute of Physics Conference Series Vol. 924, The
  Multicolored Landscape of Compact Objects and Their Explosive Origins. pp
  342--349, \mn@doi{10.1063/1.2774879}

\bibitem[\protect\citeauthoryear{{Colgate}, {Petschek}  \& {Kriese}}{{Colgate}
  et~al.}{1980}]{1980ApJ...237L..81C}
{Colgate} S.~A.,  {Petschek} A.~G.,   {Kriese} J.~T.,  1980, \mn@doi [\apjl]
  {10.1086/183239}, \href {http://adsabs.harvard.edu/abs/1980ApJ...237L..81C}
  {237, L81}

\bibitem[\protect\citeauthoryear{{Contardo}, {Leibundgut}  \&
  {Vacca}}{{Contardo} et~al.}{2000}]{2000A&A...359..876C}
{Contardo} G.,  {Leibundgut} B.,   {Vacca} W.~D.,  2000, \aap, \href
  {http://adsabs.harvard.edu/abs/2000A%26A...359..876C} {359, 876}

\bibitem[\protect\citeauthoryear{{Filippenko}, {Foley}, {Silverman}, {Blondin},
  {Matheson}  \& {Berlind}}{{Filippenko} et~al.}{2007a}]{2007CBET..817....1F}
{Filippenko} A.~V.,  {Foley} R.~J.,  {Silverman} J.~M.,  {Blondin} S.,
  {Matheson} T.,   {Berlind} P.,  2007a, Central Bureau Electronic Telegrams,
  \href {http://adsabs.harvard.edu/abs/2007CBET..817....1F} {817}

\bibitem[\protect\citeauthoryear{{Filippenko}, {Foley}, {Silverman},
  {Chornock}, {Li}, {Blondin}  \& {Matheson}}{{Filippenko}
  et~al.}{2007b}]{2007CBET..926....1F}
{Filippenko} A.~V.,  {Foley} R.~J.,  {Silverman} J.~M.,  {Chornock} R.,  {Li}
  W.,  {Blondin} S.,   {Matheson} T.,  2007b, Central Bureau Electronic
  Telegrams, \href {http://adsabs.harvard.edu/abs/2007CBET..926....1F} {926}

\bibitem[\protect\citeauthoryear{{Fink} et~al.,}{{Fink}
  et~al.}{2014}]{2014MNRAS.438.1762F}
{Fink} M.,  et~al., 2014, \mn@doi [\mnras] {10.1093/mnras/stt2315}, \href
  {http://adsabs.harvard.edu/abs/2014MNRAS.438.1762F} {438, 1762}

\bibitem[\protect\citeauthoryear{{Foley} et~al.,}{{Foley}
  et~al.}{2009}]{2009AJ....138..376F}
{Foley} R.~J.,  et~al., 2009, \mn@doi [\aj] {10.1088/0004-6256/138/2/376},
  \href {http://adsabs.harvard.edu/abs/2009AJ....138..376F} {138, 376}

\bibitem[\protect\citeauthoryear{{Foley} et~al.,}{{Foley}
  et~al.}{2010a}]{2010AJ....140.1321F}
{Foley} R.~J.,  et~al., 2010a, \mn@doi [\aj] {10.1088/0004-6256/140/5/1321},
  \href {http://cdsads.u-strasbg.fr/abs/2010AJ....140.1321F} {140, 1321}

\bibitem[\protect\citeauthoryear{{Foley}, {Brown}, {Rest}, {Challis},
  {Kirshner}  \& {Wood-Vasey}}{{Foley} et~al.}{2010b}]{2010ApJ...708L..61F}
{Foley} R.~J.,  {Brown} P.~J.,  {Rest} A.,  {Challis} P.~J.,  {Kirshner} R.~P.,
    {Wood-Vasey} W.~M.,  2010b, \mn@doi [\apjl] {10.1088/2041-8205/708/1/L61},
  \href {http://adsabs.harvard.edu/abs/2010ApJ...708L..61F} {708, L61}

\bibitem[\protect\citeauthoryear{{Foley} et~al.,}{{Foley}
  et~al.}{2013}]{2013ApJ...767...57F}
{Foley} R.~J.,  et~al., 2013, \mn@doi [\apj] {10.1088/0004-637X/767/1/57},
  \href {http://adsabs.harvard.edu/abs/2013ApJ...767...57F} {767, 57}

\bibitem[\protect\citeauthoryear{{Foley}, {Van Dyk}, {Jha}, {Clubb},
  {Filippenko}, {Mauerhan}, {Miller}  \& {Smith}}{{Foley}
  et~al.}{2015}]{2015ApJ...798L..37F}
{Foley} R.~J.,  {Van Dyk} S.~D.,  {Jha} S.~W.,  {Clubb} K.~I.,  {Filippenko}
  A.~V.,  {Mauerhan} J.~C.,  {Miller} A.~A.,   {Smith} N.,  2015, \mn@doi
  [\apjl] {10.1088/2041-8205/798/2/L37}, \href
  {http://adsabs.harvard.edu/abs/2015ApJ...798L..37F} {798, L37}

\bibitem[\protect\citeauthoryear{{Foley}, {Jha}, {Pan}, {Zheng}, {Bildsten},
  {Filippenko}  \& {Kasen}}{{Foley} et~al.}{2016}]{2016MNRAS.461..433F}
{Foley} R.~J.,  {Jha} S.~W.,  {Pan} Y.-C.,  {Zheng} W.~K.,  {Bildsten} L.,
  {Filippenko} A.~V.,   {Kasen} D.,  2016, \mn@doi [\mnras]
  {10.1093/mnras/stw1320}, \href
  {http://adsabs.harvard.edu/abs/2016MNRAS.461..433F} {461, 433}

\bibitem[\protect\citeauthoryear{{Fox} et~al.,}{{Fox}
  et~al.}{2016}]{2016ApJ...816L..13F}
{Fox} O.~D.,  et~al., 2016, \mn@doi [\apjl] {10.3847/2041-8205/816/1/L13},
  \href {http://adsabs.harvard.edu/abs/2016ApJ...816L..13F} {816, L13}

\bibitem[\protect\citeauthoryear{{Ganeshalingam} et~al.,}{{Ganeshalingam}
  et~al.}{2010}]{2010ApJS..190..418G}
{Ganeshalingam} M.,  et~al., 2010, \mn@doi [\apjs]
  {10.1088/0067-0049/190/2/418}, \href
  {http://adsabs.harvard.edu/abs/2010ApJS..190..418G} {190, 418}

\bibitem[\protect\citeauthoryear{{Gibson} \& {Stetson}}{{Gibson} \&
  {Stetson}}{2001}]{2001ApJ...547L.103G}
{Gibson} B.~K.,  {Stetson} P.~B.,  2001, \mn@doi [\apjl] {10.1086/318905},
  \href {http://adsabs.harvard.edu/abs/2001ApJ...547L.103G} {547, L103}

\bibitem[\protect\citeauthoryear{{Jha} et~al.,}{{Jha}
  et~al.}{2006}]{2006AJ....131..527J}
{Jha} S.,  et~al., 2006, \mn@doi [\aj] {10.1086/497989}, \href
  {http://cdsads.u-strasbg.fr/abs/2006AJ....131..527J} {131, 527}

\bibitem[\protect\citeauthoryear{{Jha}, {Riess}  \& {Kirshner}}{{Jha}
  et~al.}{2007}]{2007ApJ...659..122J}
{Jha} S.,  {Riess} A.~G.,   {Kirshner} R.~P.,  2007, \mn@doi [\apj]
  {10.1086/512054}, \href {http://adsabs.harvard.edu/abs/2007ApJ...659..122J}
  {659, 122}

\bibitem[\protect\citeauthoryear{{Jordan}, {Perets}, {Fisher}  \& {van
  Rossum}}{{Jordan} et~al.}{2012}]{2012ApJ...761L..23J}
{Jordan} IV G.~C.,  {Perets} H.~B.,  {Fisher} R.~T.,   {van Rossum} D.~R.,
  2012, \mn@doi [\apjl] {10.1088/2041-8205/761/2/L23}, \href
  {http://adsabs.harvard.edu/abs/2012ApJ...761L..23J} {761, L23}

\bibitem[\protect\citeauthoryear{{Kromer} et~al.,}{{Kromer}
  et~al.}{2013}]{2013MNRAS.429.2287K}
{Kromer} M.,  et~al., 2013, \mn@doi [\mnras] {10.1093/mnras/sts498}, \href
  {http://adsabs.harvard.edu/abs/2013MNRAS.429.2287K} {429, 2287}

\bibitem[\protect\citeauthoryear{{Kromer} et~al.,}{{Kromer}
  et~al.}{2015}]{2015MNRAS.450.3045K}
{Kromer} M.,  et~al., 2015, \mn@doi [\mnras] {10.1093/mnras/stv886}, \href
  {http://adsabs.harvard.edu/abs/2015MNRAS.450.3045K} {450, 3045}

\bibitem[\protect\citeauthoryear{{Landolt}}{{Landolt}}{2009}]{2009AJ....137.4186L}
{Landolt} A.~U.,  2009, \mn@doi [\aj] {10.1088/0004-6256/137/5/4186}, \href
  {http://adsabs.harvard.edu/abs/2009AJ....137.4186L} {137, 4186}

\bibitem[\protect\citeauthoryear{{Lee}, {Li}, {Newton}  \& {Puckett}}{{Lee}
  et~al.}{2007}]{2007CBET..809....1L}
{Lee} N.,  {Li} W.,  {Newton} J.,   {Puckett} T.,  2007, Central Bureau
  Electronic Telegrams, \href
  {http://adsabs.harvard.edu/abs/2007CBET..809....1L} {809}

\bibitem[\protect\citeauthoryear{{Li} et~al.,}{{Li}
  et~al.}{2001}]{2001PASP..113.1178L}
{Li} W.,  et~al., 2001, \mn@doi [\pasp] {10.1086/323355}, \href
  {http://adsabs.harvard.edu/abs/2001PASP..113.1178L} {113, 1178}

\bibitem[\protect\citeauthoryear{{Li} et~al.,}{{Li}
  et~al.}{2003}]{2003PASP..115..453L}
{Li} W.,  et~al., 2003, \mn@doi [\pasp] {10.1086/374200}, \href
  {http://adsabs.harvard.edu/abs/2003PASP..115..453L} {115, 453}

\bibitem[\protect\citeauthoryear{{Liu}, {Moriya}, {Stancliffe}  \&
  {Wang}}{{Liu} et~al.}{2015a}]{2015A&A...574A..12L}
{Liu} Z.-W.,  {Moriya} T.~J.,  {Stancliffe} R.~J.,   {Wang} B.,  2015a, \mn@doi
  [\aap] {10.1051/0004-6361/201424532}, \href
  {http://adsabs.harvard.edu/abs/2015A%26A...574A..12L} {574, A12}

\bibitem[\protect\citeauthoryear{{Liu}, {Stancliffe}, {Abate}  \& {Wang}}{{Liu}
  et~al.}{2015b}]{2015ApJ...808..138L}
{Liu} Z.-W.,  {Stancliffe} R.~J.,  {Abate} C.,   {Wang} B.,  2015b, \mn@doi
  [\apj] {10.1088/0004-637X/808/2/138}, \href
  {http://adsabs.harvard.edu/abs/2015ApJ...808..138L} {808, 138}

\bibitem[\protect\citeauthoryear{{Maeda}, {Mazzali}, {Deng}, {Nomoto},
  {Yoshii}, {Tomita}  \& {Kobayashi}}{{Maeda}
  et~al.}{2003}]{2003ApJ...593..931M}
{Maeda} K.,  {Mazzali} P.~A.,  {Deng} J.,  {Nomoto} K.,  {Yoshii} Y.,  {Tomita}
  H.,   {Kobayashi} Y.,  2003, \mn@doi [\apj] {10.1086/376591}, \href
  {http://adsabs.harvard.edu/abs/2003ApJ...593..931M} {593, 931}

\bibitem[\protect\citeauthoryear{{Magee} et~al.,}{{Magee}
  et~al.}{2016}]{2016A&A...589A..89M}
{Magee} M.~R.,  et~al., 2016, \mn@doi [\aap] {10.1051/0004-6361/201528036},
  \href {http://adsabs.harvard.edu/abs/2016A%26A...589A..89M} {589, A89}

\bibitem[\protect\citeauthoryear{{McClelland} et~al.,}{{McClelland}
  et~al.}{2010}]{2010ApJ...720..704M}
{McClelland} C.~M.,  et~al., 2010, \mn@doi [\apj]
  {10.1088/0004-637X/720/1/704}, \href
  {http://adsabs.harvard.edu/abs/2010ApJ...720..704M} {720, 704}

\bibitem[\protect\citeauthoryear{{McCully} et~al.,}{{McCully}
  et~al.}{2014a}]{2014Natur.512...54M}
{McCully} C.,  et~al., 2014a, \mn@doi [\nat] {10.1038/nature13615}, \href
  {http://adsabs.harvard.edu/abs/2014Natur.512...54M} {512, 54}

\bibitem[\protect\citeauthoryear{{McCully} et~al.,}{{McCully}
  et~al.}{2014b}]{2014ApJ...786..134M}
{McCully} C.,  et~al., 2014b, \mn@doi [\apj] {10.1088/0004-637X/786/2/134},
  \href {http://adsabs.harvard.edu/abs/2014ApJ...786..134M} {786, 134}

\bibitem[\protect\citeauthoryear{{Milne}, {The}  \& {Leising}}{{Milne}
  et~al.}{1999}]{1999ApJS..124..503M}
{Milne} P.~A.,  {The} L.-S.,   {Leising} M.~D.,  1999, \mn@doi [\apjs]
  {10.1086/313262}, \href {http://adsabs.harvard.edu/abs/1999ApJS..124..503M}
  {124, 503}

\bibitem[\protect\citeauthoryear{{Milne}, {The}  \& {Leising}}{{Milne}
  et~al.}{2001}]{2001ApJ...559.1019M}
{Milne} P.~A.,  {The} L.-S.,   {Leising} M.~D.,  2001, \mn@doi [\apj]
  {10.1086/322352}, \href {http://adsabs.harvard.edu/abs/2001ApJ...559.1019M}
  {559, 1019}

\bibitem[\protect\citeauthoryear{{Moriya}, {Tominaga}, {Tanaka}, {Nomoto},
  {Sauer}, {Mazzali}, {Maeda}  \& {Suzuki}}{{Moriya}
  et~al.}{2010}]{2010ApJ...719.1445M}
{Moriya} T.,  {Tominaga} N.,  {Tanaka} M.,  {Nomoto} K.,  {Sauer} D.~N.,
  {Mazzali} P.~A.,  {Maeda} K.,   {Suzuki} T.,  2010, \mn@doi [\apj]
  {10.1088/0004-637X/719/2/1445}, \href
  {http://adsabs.harvard.edu/abs/2010ApJ...719.1445M} {719, 1445}

\bibitem[\protect\citeauthoryear{{Nakano} et~al.,}{{Nakano}
  et~al.}{2014}]{2014CBET.4011....1N}
{Nakano} S.,  et~al., 2014, Central Bureau Electronic Telegrams, \href
  {http://adsabs.harvard.edu/abs/2014CBET.4011....1N} {4011}

\bibitem[\protect\citeauthoryear{{Narayan} et~al.,}{{Narayan}
  et~al.}{2011}]{2011ApJ...731L..11N}
{Narayan} G.,  et~al., 2011, \mn@doi [\apjl] {10.1088/2041-8205/731/1/L11},
  \href {http://adsabs.harvard.edu/abs/2011ApJ...731L..11N} {731, L11}

\bibitem[\protect\citeauthoryear{{Ochner}, {Tomasella}, {Benetti},
  {Cappellaro}, {Elias-Rosa}, {Pastorello}  \& {Turatto}}{{Ochner}
  et~al.}{2014}]{2014CBET.4011....2O}
{Ochner} P.,  {Tomasella} L.,  {Benetti} S.,  {Cappellaro} E.,  {Elias-Rosa}
  N.,  {Pastorello} A.,   {Turatto} M.,  2014, Central Bureau Electronic
  Telegrams, \href {http://adsabs.harvard.edu/abs/2014CBET.4011....2O} {4011}

\bibitem[\protect\citeauthoryear{{Pejcha} \& {Prieto}}{{Pejcha} \&
  {Prieto}}{2015}]{2015ApJ...799..215P}
{Pejcha} O.,  {Prieto} J.~L.,  2015, \mn@doi [\apj]
  {10.1088/0004-637X/799/2/215}, \href
  {http://adsabs.harvard.edu/abs/2015ApJ...799..215P} {799, 215}

\bibitem[\protect\citeauthoryear{{Phillips} et~al.,}{{Phillips}
  et~al.}{2007}]{2007PASP..119..360P}
{Phillips} M.~M.,  et~al., 2007, \mn@doi [\pasp] {10.1086/518372}, \href
  {http://adsabs.harvard.edu/abs/2007PASP..119..360P} {119, 360}

\bibitem[\protect\citeauthoryear{{Prabhu} \& {Anupama}}{{Prabhu} \&
  {Anupama}}{2010}]{2010ASInC...1..193P}
{Prabhu} T.~P.,  {Anupama} G.~C.,  2010, in Astronomical Society of India
  Conference Series.

\bibitem[\protect\citeauthoryear{{Rajala} et~al.,}{{Rajala}
  et~al.}{2005}]{2005PASP..117..132R}
{Rajala} A.~M.,  et~al., 2005, \mn@doi [\pasp] {10.1086/427985}, \href
  {http://adsabs.harvard.edu/abs/2005PASP..117..132R} {117, 132}

\bibitem[\protect\citeauthoryear{{Rodr{\'{\i}}guez}, {Clocchiatti}  \&
  {Hamuy}}{{Rodr{\'{\i}}guez} et~al.}{2014}]{2014AJ....148..107R}
{Rodr{\'{\i}}guez} {\'O}.,  {Clocchiatti} A.,   {Hamuy} M.,  2014, \mn@doi
  [\aj] {10.1088/0004-6256/148/6/107}, \href
  {http://adsabs.harvard.edu/abs/2014AJ....148..107R} {148, 107}

\bibitem[\protect\citeauthoryear{{Roy} et~al.,}{{Roy}
  et~al.}{2011}]{2011ApJ...736...76R}
{Roy} R.,  et~al., 2011, \mn@doi [\apj] {10.1088/0004-637X/736/2/76}, \href
  {http://adsabs.harvard.edu/abs/2011ApJ...736...76R} {736, 76}

\bibitem[\protect\citeauthoryear{{Ruiz-Lapuente} \& {Spruit}}{{Ruiz-Lapuente}
  \& {Spruit}}{1998}]{1998ApJ...500..360R}
{Ruiz-Lapuente} P.,  {Spruit} H.~C.,  1998, \mn@doi [\apj] {10.1086/305697},
  \href {http://adsabs.harvard.edu/abs/1998ApJ...500..360R} {500, 360}

\bibitem[\protect\citeauthoryear{{Sagar}}{{Sagar}}{1999}]{1999CSci...77..643G}
{Sagar} R.,  1999, Current Science, \href
  {http://adsabs.harvard.edu/abs/1999CSci...77..643G} {77, 643}

\bibitem[\protect\citeauthoryear{{Sagar}, {Kumar}, {Omar}  \& {Pandey}}{{Sagar}
  et~al.}{2012}]{2012SPIE.8444E..1TS}
{Sagar} R.,  {Kumar} B.,  {Omar} A.,   {Pandey} A.~K.,  2012, in Ground-based
  and Airborne Telescopes IV. p. 84441T (\mn@eprint {arXiv} {1304.2474}),
  \mn@doi{10.1117/12.925634}

\bibitem[\protect\citeauthoryear{{Sahu} et~al.,}{{Sahu}
  et~al.}{2008}]{2008ApJ...680..580S}
{Sahu} D.~K.,  et~al., 2008, \mn@doi [\apj] {10.1086/587772}, \href
  {http://adsabs.harvard.edu/abs/2008ApJ...680..580S} {680, 580}

\bibitem[\protect\citeauthoryear{{Schlafly} \& {Finkbeiner}}{{Schlafly} \&
  {Finkbeiner}}{2011}]{2011ApJ...737..103S}
{Schlafly} E.~F.,  {Finkbeiner} D.~P.,  2011, \mn@doi [\apj]
  {10.1088/0004-637X/737/2/103}, \href
  {http://adsabs.harvard.edu/abs/2011ApJ...737..103S} {737, 103}

\bibitem[\protect\citeauthoryear{{Schoeniger} \& {Sofue}}{{Schoeniger} \&
  {Sofue}}{1997}]{1997A&A...323...14S}
{Schoeniger} F.,  {Sofue} Y.,  1997, \aap, \href
  {http://adsabs.harvard.edu/abs/1997A%26A...323...14S} {323, 14}

\bibitem[\protect\citeauthoryear{{Sparks}}{{Sparks}}{1994}]{1994ApJ...433...19S}
{Sparks} W.~B.,  1994, \mn@doi [\apj] {10.1086/174621}, \href
  {http://adsabs.harvard.edu/abs/1994ApJ...433...19S} {433, 19}

\bibitem[\protect\citeauthoryear{{Stalin}, {Hegde}, {Sahu}, {Parihar},
  {Anupama}, {Bhatt}  \& {Prabhu}}{{Stalin} et~al.}{2008}]{2008BASI...36..111S}
{Stalin} C.~S.,  {Hegde} M.,  {Sahu} D.~K.,  {Parihar} P.~S.,  {Anupama} G.~C.,
   {Bhatt} B.~C.,   {Prabhu} T.~P.,  2008, Bulletin of the Astronomical Society
  of India, \href {http://adsabs.harvard.edu/abs/2008BASI...36..111S} {36, 111}

\bibitem[\protect\citeauthoryear{{Stetson}}{{Stetson}}{1987}]{1987PASP...99..191S}
{Stetson} P.~B.,  1987, \mn@doi [\pasp] {10.1086/131977}, \href
  {http://adsabs.harvard.edu/abs/1987PASP...99..191S} {99, 191}

\bibitem[\protect\citeauthoryear{{Stetson}}{{Stetson}}{1992}]{1992JRASC..86...71S}
{Stetson} P.~B.,  1992, \jrasc, \href
  {http://adsabs.harvard.edu/abs/1992JRASC..86...71S} {86, 71}

\bibitem[\protect\citeauthoryear{{Stritzinger} \& {Sollerman}}{{Stritzinger} \&
  {Sollerman}}{2007}]{2007A&A...470L...1S}
{Stritzinger} M.,  {Sollerman} J.,  2007, \mn@doi [\aap]
  {10.1051/0004-6361:20066999}, \href
  {http://adsabs.harvard.edu/abs/2007A%26A...470L...1S} {470, L1}

\bibitem[\protect\citeauthoryear{{Stritzinger} et~al.,}{{Stritzinger}
  et~al.}{2014}]{2014A&A...561A.146S}
{Stritzinger} M.~D.,  et~al., 2014, \mn@doi [\aap]
  {10.1051/0004-6361/201322889}, \href
  {http://adsabs.harvard.edu/abs/2014A%26A...561A.146S} {561, A146}

\bibitem[\protect\citeauthoryear{{Stritzinger} et~al.,}{{Stritzinger}
  et~al.}{2015}]{2015A&A...573A...2S}
{Stritzinger} M.~D.,  et~al., 2015, \mn@doi [\aap]
  {10.1051/0004-6361/201424168}, \href
  {http://adsabs.harvard.edu/abs/2015A%26A...573A...2S} {573, A2}

\bibitem[\protect\citeauthoryear{Suntzeff}{Suntzeff}{1996}]{sent:1996}
Suntzeff N.~B.,  1996, in McCray R.,  Wang Z.,  eds, , Supernovae and Supernova
  Remnants.
Cambridge University Press, Cambridge, p.~41

\bibitem[\protect\citeauthoryear{{Szalai} et~al.,}{{Szalai}
  et~al.}{2015}]{2015MNRAS.453.2103S}
{Szalai} T.,  et~al., 2015, \mn@doi [\mnras] {10.1093/mnras/stv1776}, \href
  {http://adsabs.harvard.edu/abs/2015MNRAS.453.2103S} {453, 2103}

\bibitem[\protect\citeauthoryear{{Thomas}, {Nugent}  \& {Meza}}{{Thomas}
  et~al.}{2011}]{2011PASP..123..237T}
{Thomas} R.~C.,  {Nugent} P.~E.,   {Meza} J.~C.,  2011, \mn@doi [\pasp]
  {10.1086/658673}, \href {http://adsabs.harvard.edu/abs/2011PASP..123..237T}
  {123, 237}

\bibitem[\protect\citeauthoryear{{Tomasella} et~al.,}{{Tomasella}
  et~al.}{2016}]{2016MNRAS.459.1018T}
{Tomasella} L.,  et~al., 2016, \mn@doi [\mnras] {10.1093/mnras/stw696}, \href
  {http://adsabs.harvard.edu/abs/2016MNRAS.459.1018T} {459, 1018}

\bibitem[\protect\citeauthoryear{{Turatto}, {Benetti}, {Cappellaro},
  {Danziger}, {Della Valle}, {Gouiffes}, {Mazzali}  \& {Patat}}{{Turatto}
  et~al.}{1996}]{1996MNRAS.283....1T}
{Turatto} M.,  {Benetti} S.,  {Cappellaro} E.,  {Danziger} I.~J.,  {Della
  Valle} M.,  {Gouiffes} C.,  {Mazzali} P.~A.,   {Patat} F.,  1996, \mn@doi
  [\mnras] {10.1093/mnras/283.1.1}, \href
  {http://adsabs.harvard.edu/abs/1996MNRAS.283....1T} {283, 1}

\bibitem[\protect\citeauthoryear{{Valenti} et~al.,}{{Valenti}
  et~al.}{2009}]{2009Natur.459..674V}
{Valenti} S.,  et~al., 2009, \mn@doi [\nat] {10.1038/nature08023}, \href
  {http://adsabs.harvard.edu/abs/2009Natur.459..674V} {459, 674}

\bibitem[\protect\citeauthoryear{{White} et~al.,}{{White}
  et~al.}{2015}]{2015ApJ...799...52W}
{White} C.~J.,  et~al., 2015, \mn@doi [\apj] {10.1088/0004-637X/799/1/52},
  \href {http://cdsads.u-strasbg.fr/abs/2015ApJ...799...52W} {799, 52}

\bibitem[\protect\citeauthoryear{{Willcox}, {Townsley}, {Calder}, {Denissenkov}
   \& {Herwig}}{{Willcox} et~al.}{2016}]{2016ApJ...832...13W}
{Willcox} D.~E.,  {Townsley} D.~M.,  {Calder} A.~C.,  {Denissenkov} P.~A.,
  {Herwig} F.,  2016, \mn@doi [\apj] {10.3847/0004-637X/832/1/13}, \href
  {http://adsabs.harvard.edu/abs/2016ApJ...832...13W} {832, 13}

\bibitem[\protect\citeauthoryear{{Yamanaka} et~al.,}{{Yamanaka}
  et~al.}{2015}]{2015ApJ...806..191Y}
{Yamanaka} M.,  et~al., 2015, \mn@doi [\apj] {10.1088/0004-637X/806/2/191},
  \href {http://adsabs.harvard.edu/abs/2015ApJ...806..191Y} {806, 191}

\bibitem[\protect\citeauthoryear{{Yaron} \& {Gal-Yam}}{{Yaron} \&
  {Gal-Yam}}{2012}]{2012PASP..124..668Y}
{Yaron} O.,  {Gal-Yam} A.,  2012, \mn@doi [\pasp] {10.1086/666656}, \href
  {http://adsabs.harvard.edu/abs/2012PASP..124..668Y} {124, 668}

\bibitem[\protect\citeauthoryear{{van Dokkum}}{{van
  Dokkum}}{2001}]{2001PASP..113.1420V}
{van Dokkum} P.~G.,  2001, \mn@doi [\pasp] {10.1086/323894}, \href
  {http://adsabs.harvard.edu/abs/2001PASP..113.1420V} {113, 1420}

\makeatother
\end{thebibliography}
\begin{table*}
\caption{ Details of SN 2014dt and it's host galaxy M61 }
\centering
\smallskip
%\footnote
\begin{tabular}{c c }
\hline \hline
Host galaxy$^\star$ & M61    \\
Galaxy type & SAB(rs)bc \\  
Constellation & Virgo  \\ 
Redshift & 0.005224$\pm$0.000007 \\ 
Major Diameter & 6.5 arcmin \\
Minor Diameter & 5.8 arcmin \\
Helio. Radial Velocity &  1566$\pm$2 km/sec \\
\hline
R.A.(J2000.0) & 12$^h$21$^m$57.57$^s$ \\
Dec.(J2000.0) & +04$^d$28$^m$18.5$^s$ \\
Luminosity Distance  & 21.44$\pm$0.03 Mpc \\
Galactic Extincion E(B-V) & 0.02 mag \\
SN type & Iax\\
Offset from nucleus & 33$^{''}$.9 E,7$^{''}$.2 S \\
Date of Discovery & 2456960.338 (JD) \\
\hline 
\end{tabular}
\newline
$^\star$ Host galaxy parameters taken from NED                     
\label{tab:sn14dt_m61_detail}      
\end{table*}

\begin{table*}
\caption{ Details of Instruments and Detectors }
\centering
\smallskip
%\footnote
\begin{tabular}{c c c c c c c c}
\hline \hline
Telescope & Detector      & area              & Pixel Size     & Plate Scale      & Read out noise & Gain        \\
          &               & (arcminute$^2$)   & ($\micron$)    &(arcsecond/pixel) & (e$^-$)        & (e$^-$/ADU) \\
\hline
ST        & 2k$\times$2k  & 13.5$\times$13.5  & 24             & 0.37               &5.3             &10.00       \\
ST        & 1k$\times$1k  & 6.5$\times$6.5   & 24             & 0.37               &7.0             &11.98       \\
DFOT      & 2k$\times$2k  & 18.0$\times$18.0  & 13.5           & 0.54               &7.0             &2.0          \\
HCT       & 2k$\times$4k  & 10.0$\times$10.0  & 15             & 0.17               &4.8             &1.22         \\
\hline                                   
\end{tabular}
\label{tab:details_instrument_detectors}      
\end{table*}

\begin{table*}
\caption{Star ID along with magnitude of secondary standard stars used for calibration}
\centering
\smallskip
\begin{tabular}{c c c c c c c }
\hline \hline
Star ID  & R.A.(h:m:s) & Dec.(d:m:s)      &  B(mag)              &       V(mag)                  &      R(mag)                   &      I(mag)               \\
                                      
\hline
A        & 12:22:07.92  & +04:33:04.5     &  16.984 $\pm$ 0.028	 & 	16.728 $\pm$ 0.011 	 & 	16.411 $\pm$  	0.010	 & 	16.135  $\pm$   0.011 \\
B        & 12:22:04.10  & +04:32:29.5     &  15.997 $\pm$ 0.027	 & 	15.438 $\pm$ 0.011 	 & 	14.971 $\pm$  	0.009	 & 	14.606  $\pm$   0.008 \\
C        & 12:22:06.00  & +04:30:52.1     &  17.943 $\pm$ 0.031	 & 	17.579 $\pm$ 0.017 	 & 	17.246 $\pm$  	0.010	 & 	16.976  $\pm$   0.017 \\
D        & 12:22:02.37  & +04:30:08.6     &  19.089 $\pm$ 0.051	 & 	17.944 $\pm$ 0.013 	 & 	17.023 $\pm$  	0.010	 & 	16.212  $\pm$   0.014 \\
E        & 12:22:06.08  & +04:28:41.4     &  18.419 $\pm$ 0.036	 & 	17.429 $\pm$ 0.012 	 & 	16.650 $\pm$  	0.010	 & 	16.024  $\pm$   0.010 \\
F        & 12:22:02.96  & +04:24:52.2     &  17.923 $\pm$ 0.031	 & 	16.910 $\pm$ 0.011 	 & 	16.105 $\pm$  	0.009	 & 	15.481  $\pm$   0.009 \\
G        & 12:22:06.76  & +04:23:23.5     &  17.916 $\pm$ 0.031	 & 	16.841 $\pm$ 0.011 	 & 	16.046 $\pm$  	0.009	 & 	15.404  $\pm$   0.009 \\
H        & 12:21:56.37  & +04:23:49.3     &  15.090 $\pm$ 0.027	 & 	14.647 $\pm$ 0.011 	 & 	14.271 $\pm$  	0.009	 & 	13.951  $\pm$   0.008 \\
I        & 12:21:44.94  & +04:23:46.0     &  17.309 $\pm$ 0.029	 & 	16.724 $\pm$ 0.011 	 & 	16.221 $\pm$  	0.009	 & 	15.815  $\pm$   0.010 \\
J        & 12:21:43.86  & +04:24:23.3     &  17.596 $\pm$ 0.030	 & 	17.235 $\pm$ 0.012 	 & 	16.894 $\pm$  	0.010	 & 	16.618  $\pm$   0.011 \\
K        & 12:21:44.50  & +04:25:23.9     &  17.771 $\pm$ 0.030	 & 	17.488 $\pm$ 0.012 	 & 	17.174 $\pm$  	0.011	 & 	16.876  $\pm$   0.014 \\
L        & 12:21:48.98  & +04:26:20.8     &  14.237 $\pm$ 0.027	 & 	13.872 $\pm$ 0.011 	 & 	13.529 $\pm$  	0.009	 & 	13.324  $\pm$   0.011 \\
M        & 12:21:40.25  & +04:28:17.4     &  18.915 $\pm$ 0.044	 & 	16.924 $\pm$ 0.012 	 & 	16.635 $\pm$  	0.013	 & 	14.993  $\pm$   0.009 \\
N        & 12:21:39.45  & +04:29:04.8     &  17.555 $\pm$ 0.029	 & 	17.095 $\pm$ 0.012 	 & 	16.660 $\pm$  	0.010	 & 	16.283  $\pm$   0.010 \\
O        & 12:21:37.95  & +04:30:26.5     &  17.482 $\pm$ 0.029	 & 	17.211 $\pm$ 0.014 	 & 	16.804 $\pm$  	0.013	 & 	16.272  $\pm$   0.020 \\
P        & 12:21:48.24  & +04:32:26.0     &  17.025 $\pm$ 0.028	 & 	15.898 $\pm$ 0.011 	 & 	14.967 $\pm$  	0.009	 & 	14.087  $\pm$   0.008 \\
Q        & 12:21:52.85  & +04:30:25.9     &  16.344 $\pm$ 0.027	 & 	15.456 $\pm$ 0.011 	 & 	14.768 $\pm$  	0.009	 & 	14.166  $\pm$   0.008 \\ 
\hline                                   
\end{tabular}
\label{tab:standard_star_table}      
\end{table*}

\begin{table*}
\caption{Log of optical observations}
\centering
\smallskip
\begin{tabular}{c c c c c c c}
\hline \hline
Date    &   Phase$^\dagger$       &   B            &   V                          &   R                       &  I                    & Telescope     \\
        &   (Days)                & (mag)          & (mag)                        & (mag)                     & (mag)                                 \\
\hline  
20141030         &  9.83    &  14.105 $\pm$ 0.015        & 13.620 $\pm$ 	0.037    &  13.296 $\pm$  0.020       &  13.255 $\pm$   0.009 & HCT \\    
20141123         &  33.17   &  16.186 $\pm$ 0.019        & 15.015 $\pm$ 	0.017    &  14.610 $\pm$  0.024       &  14.179 $\pm$   0.012 & HCT \\
20141125         &  36.14   &  16.156 $\pm$ 0.015        & 15.039 $\pm$ 	0.019    &  14.610 $\pm$  0.023       &  --                   & DFOT\\
20141126         &  37.12   &  16.161 $\pm$ 0.014        & 15.061 $\pm$         0.019   &  14.626 $\pm$  0.025       &  14.238 $\pm$   0.013 & DFOT \\
20141201         &  42.11   &  16.338 $\pm$ 0.014        & 15.221 $\pm$         0.016   &  14.940 $\pm$  0.023       &  14.447 $\pm$	0.011 & HCT\\
20141204         &  45.13   &  16.414 $\pm$ 0.015        & 15.336 $\pm$ 	0.017    &  14.961 $\pm$  0.023       &  14.556 $\pm$	0.010 & HCT\\
20141205         &  46.38   &  16.196 $\pm$ 0.024        & 15.309 $\pm$ 	0.019    &  14.973 $\pm$  0.024       &  14.573 $\pm$	0.017 & ST\\
20141208         &  49.15   &  16.418 $\pm$ 0.023        & 15.427 $\pm$ 	0.018    &  15.059 $\pm$  0.030       &  14.642 $\pm$	0.024 & HCT\\
20141209         &  50.38   &  16.524 $\pm$ 0.025        &   --                         &  15.069 $\pm$  0.024       &  14.675 $\pm$	0.015 & ST\\
20141210         &  51.33   &  16.462 $\pm$ 0.028        & 15.467 $\pm$ 	0.021    &  15.090 $\pm$  0.024       &  14.701 $\pm$	0.017 & ST\\
20141216         &  57.38   &      --                    & 15.618 $\pm$ 	0.019    &  15.264 $\pm$  0.024       &  14.876 $\pm$	0.012 & ST\\
20141229         &  70.33   &      --                    & 15.893 $\pm$ 	0.028    &  15.651 $\pm$  0.028       &  15.187 $\pm$   0.016 & ST\\
20150104         &  76.32   &      --                    & 16.095 $\pm$ 	0.018    &  15.702 $\pm$  0.025       &  --                   & ST\\
20150106         &  78.39   &      --                    & 16.216 $\pm$ 	0.026    &  15.767 $\pm$  0.027       &  15.336 $\pm$ 	0.019 & ST\\  
20150107         &  78.99   &      --                    & 16.131 $\pm$ 	0.022    &  15.773 $\pm$  0.024       &  15.324 $\pm$ 	0.017 & HCT\\          
20150109         &  81.29   &      --                    & 16.094 $\pm$ 	0.020    &  15.800 $\pm$  0.025       &  15.375 $\pm$ 	0.013 & ST\\            
20150120         &  92.29   &      --                    & 16.303 $\pm$ 	0.024    &  15.986 $\pm$  0.028       &  15.556 $\pm$ 	0.017 & ST\\           
20150130         &  102.28  &       --                   & 16.479 $\pm$ 	0.023    &  16.153 $\pm$  0.029       &  15.668 $\pm$ 	0.021 & ST\\           
20150204         &  107.36  &   17.489$\pm$ 0.097        & 16.652 $\pm$ 	0.035    &  16.226 $\pm$  0.028       &  15.774 $\pm$ 	0.022 & ST\\     
20150406         &  168.17  &       --                   & 17.313 $\pm$ 	0.032    &  16.729 $\pm$  0.028       &  16.189 $\pm$ 	0.018 & ST\\     
20150407         &  169.16  &   18.424$\pm$ 0.061        & 17.296 $\pm$ 	0.025    &  16.775 $\pm$  0.031       &  16.233 $\pm$ 	0.024 & ST\\     
20150408         &  170.12  &   18.177$\pm$0.053         & 17.257 $\pm$ 	0.031    &  16.894 $\pm$  0.029       &  16.272 $\pm$   0.023 & ST\\     
20150418         &  180.11  &       --                   & 17.384 $\pm$ 	0.032    &  16.855 $\pm$  0.029       &  16.216 $\pm$ 	0.028 & ST\\  
20150421         &  183.12  &       --                   & 17.412 $\pm$ 	0.034    &  16.850 $\pm$  0.032       &  16.216 $\pm$ 	0.025 & ST\\     
20150424         &  186.15  &       --                   & 17.435 $\pm$ 	0.030    &  16.859 $\pm$  0.033       &  16.371 $\pm$ 	0.029 & ST\\     
20150502         &  194.13  &       --                   & 17.638 $\pm$ 	0.062    &  16.940 $\pm$  0.039       &  16.370 $\pm$ 	0.029 & ST\\     
20150503         &  195.06  &       --                   & 17.663 $\pm$ 	0.049    &  17.100 $\pm$  0.053       &  16.369 $\pm$   0.023 & ST\\     
20150505         &  197.14  &       --                   & 17.520 $\pm$ 	0.032    &  17.007 $\pm$  0.031       &  16.189 $\pm$ 	0.044 & ST\\  
20150530         &  222.44  &        --                  & --                           &    --                      &  16.784 $\pm$ 	0.023 & ST\\     
20151201         &  407.06  &       --                   & 19.024 $\pm$ 	0.025    &  18.323 $\pm$  0.026       &  17.643 $\pm$ 	0.014 & HCT\\               
\hline    
\end{tabular}
\newline
$^\dagger$ Phase has been calculated with respect to B$_{max}$ = 2456950.34                                                                                       
\label{tab:photometric_observational_log}                                                        
\end{table*}

\begin{table*}
\caption{Log of spectroscopic observations}
\centering
\smallskip
\begin{tabular}{c c c c c}
\hline \hline
Date          & Phase$^\dagger$          & Grism      & Spectral Range        & Telescope       \\
              &(Days)                    &            & (\AA)                 &                 \\
\hline
20141111      & 23.14                    & Gr07,Gr08       & 3800-6840,5800-8350       & HFOSC, HCT  \\
20141117      & 29.16                    & Gr07,Gr08       & 3800-6840,5800-8350       & HFOSC, HCT  \\
20141123      & 35.13                    & Gr07,Gr08       & 3800-6840,5800-8350       & HFOSC, HCT  \\
20141201      & 42.16                    & Gr07,Gr08 	  & 3800-6840,5800-8350       & HFOSC, HCT  \\
20141204      & 46.09                    & Gr07,Gr08       & 3800-6840,5800-8350       & HFOSC, HCT  \\
20141208      & 49.16                    & Gr07,Gr08       & 3800-6840,5800-8350       & HFOSC, HCT  \\
20141210      & 52.1                     & Gr07            & 3800-6840                 & HFOSC, HCT   \\
20141219      & 61.07                    & Gr07,Gr08       & 3800-6840,5800-8350       & HFOSC, HCT    \\
20141226      & 67.17                    & Gr07            & 3800-6840                 & HFOSC, HCT   \\
20150104      & 77.19                    & Gr07,Gr08       & 3800-6840,5800-8350       & HFOSC, HCT    \\
20150106      & 79                       & Gr07,Gr08       & 3800-6840,5800-8350       & HFOSC, HCT     \\
20150118      & 91.09                    & Gr07,Gr08       & 3800-6840,5800-8350       & HFOSC, HCT     \\
20150125      & 97.97                    & Gr07,Gr08       & 3800-6840,5800-8350       & HFOSC, HCT     \\
20150201      & 105.02                   & Gr07,Gr08       & 3800-6840,5800-8350       & HFOSC, HCT     \\
20150205      & 109.09                   & Gr07,Gr08       & 3800-6840,5800-8350       & HFOSC, HCT     \\
20150223      & 127.02                   & Gr07,Gr08       & 3800-6840,5800-8350       & HFOSC, HCT     \\
20150312      & 143.95                   & Gr07,Gr08       & 3800-6840,5800-8350       & HFOSC, HCT    \\
20150325      & 156.89                   & Gr07,Gr08       & 3800-6840,5800-8350       & HFOSC, HCT     \\
\hline                                   
\end{tabular}
\newline
$^\dagger$ Phase has been calculated with respect to B$_{max}$= 2456950.34
\label{tab:spectroscopic_observations}      
\end{table*}

\begin{table*}
\caption{ Parameters of SN 2014dt  }
\centering
\smallskip
%\footnote
\begin{tabular}{l  c c c c}
\hline \hline
SN 2014dt                                                  & B band$^\dagger$ & V band            & R band           & I band            \\
\hline
Epoch of maximum                                           & 950.34           & 959.71            & --               & --                 \\
Magnitude at maximum (mag)                                 & 13.59$\pm$0.04 & 13.39$\pm$0.02      & --               & --                  \\
Absolute magnitude at maximum (mag)                        & -18.13$\pm$0.04 & -18.33$\pm$0.02      & --               & --                  \\
$\Delta$m$_{15}$ (mag)                                     & 1.35$\pm$0.06    & --                & --               & --   	       \\
\hline
&Decline rate mag (100 days)$^{-1}$     &   &     &    \\
\hline
Time range (days)     &   &     &     &    \\
\hline
34 -- 108 & 1.96$\pm$0.07     & 2.46$\pm$0.03     & 2.43$\pm$0.03    & 2.14$\pm$0.03    \\
168 -- 200 & --               & 1.16$\pm$0.18     & 0.82$\pm$0.21    & --    \\
168 -- 220 & --               & --                & --               & 0.87$\pm$0.22    \\
\hline
\hline
SN 2005hk &Decline rate mag (100 days)$^{-1}$    &    &     &    \\
\hline
Time range (days)     &   &     &     &    \\
\hline
20 -- 45    & 2.1              & 2.7               & 3.1              & 3.1\\
230 -- 380  & --               & 1.5               & 1.1              & --\\
\hline
\end{tabular}
\newline
$^\dagger$ JD 2,453,000+.
\label{tab:decay rate}      
\end{table*}

\begin{table*}
\caption{Properties of the comparison sample }
\centering
\smallskip
%\footnote
\begin{tabular}{c c c c c c c c }
\hline \hline
                       & SNe         & Distance      & Extinction &  M$_B$             & M$_V$            & $\Delta$m$_{15}$ &  Reference$^\dagger$  \\
                       &             & (Mpc)         & E(B-V)(mag)     &  (mag)             & (mag)            & (mag)                   \\
\hline
                       
                       & SN 2002cx   & 102.7$\pm$7.2 & 0.034      & -17.68$\pm$0.10    & -17.57$\pm$0.15  & 1.29$\pm$0.11    &  1   \\
                       & SN 2005hk   & 48.6$\pm$3.4  & 0.110      & -18.02$\pm$0.32    & -18.08$\pm$0.29  & 1.68$\pm$0.05    &  2   \\
                       & SN 2008ha   & 21.3          & 0.081      & -13.74$\pm$0.15    & -14.21$\pm$0.15  & 2.17$\pm$0.02    &  3   \\
		       & SN 2010ae$^\ddagger$   & 13.1$\pm$3.5  & 0.600      & -13.44$\pm$0.54    & -13.80$\pm$0.54  & 2.43$\pm$0.11    &  4   \\
                       & SN 2012Z{*}    & 29.8$\pm$3.8  & 0.036/0.11$\pm$0.03      & -18.27$\pm$0.09    & -18.50$\pm$0.09  & 1.43$\pm$0.02    &  5   \\
                       & SN 2014ck   & 24.4$\pm$1.7  & 0.500      & -17.37$\pm$0.15    & -17.29$\pm$0.15  & 1.76$\pm$0.15    &  6   \\
 \hline                                                                                   
\end{tabular}
\newline
$^\dagger$ REFERENCES.-- (1)\cite{2003PASP..115..453L}, NED; (2)\cite{2008ApJ...680..580S,2007PASP..119..360P}, NED; (3)\cite{2009AJ....138..376F}; (4)\cite{2014A&A...561A.146S}; (5)\cite{2015A&A...573A...2S,2015ApJ...806..191Y}; (6)\cite{2016MNRAS.459.1018T}, NED.
$^\ddagger$ A range of M$_B$ and M$_V$ is given for SN 2010ae in \cite{2014A&A...561A.146S}, we have mentioned the lower value in this paper. * Two different total extinction values were adopted by \cite{2015ApJ...806..191Y} and \cite{2015A&A...573A...2S}. For constructing bolometric light curve and colour curve of SN 2012Z we have used extinction value adopted by \cite{2015A&A...573A...2S} and data files were taken from \cite{2015ApJ...806..191Y}.        
\label{tab:photometric_parameters_different_SNe}
\end{table*}

% Don't change these lines
\bsp	% Typesetting comment
\label{lastpage}
\end{document}